\documentclass[conference]{IEEEtran}
\IEEEoverridecommandlockouts
\usepackage[margin=1in]{geometry}
\usepackage{microtype}
\usepackage{lmodern}                    
\usepackage[utf8]{inputenc}             
\usepackage{amsmath,amssymb,amsfonts,braket}
\usepackage{siunitx}
\usepackage{booktabs}
\usepackage{graphicx}
\usepackage{subcaption}
\usepackage{xspace}
\usepackage{tabularx}
\usepackage{xcolor}
\usepackage{enumitem}
\usepackage{csquotes}
\usepackage{changepage}
\usepackage{quantikz}
\usepackage{tikz}
\usepackage{algorithm}
\usepackage{algpseudocode}
\usepackage[numbers,sort&compress]{natbib}  
\usepackage{hyperref}                       
\usepackage[nameinlink,capitalise]{cleveref} 
\usepackage{bm}

\hypersetup{
  colorlinks=true,
  linkcolor=blue,
  citecolor=teal!60!black,
  urlcolor=magenta!70!black
}
\tikzset{
  node/.style = {draw, rounded corners, inner sep=6pt, font=\small},
  op/.style   = {font=\small\itshape, inner sep=0pt},
  arr/.style  = {-{Stealth}, line width=.6pt},
}

\newcommand{\qiskit}{\textsc{Qiskit}\xspace}

\DeclareUnicodeCharacter{2192}{\ifmmode\rightarrow\else{$\rightarrow$}\fi}
\DeclareUnicodeCharacter{03C0}{\ifmmode\pi\else{$\pi$}\fi}
\DeclareUnicodeCharacter{03B8}{\ifmmode\theta\else{$\theta$}\fi}
\DeclareUnicodeCharacter{2208}{\ifmmode\in\else{$\in$}\fi}
\DeclareUnicodeCharacter{00D7}{\ifmmode\times\else{$\times$}\fi}
\DeclareUnicodeCharacter{2218}{\ifmmode\circ\else{$\circ$}\fi}
\DeclareUnicodeCharacter{2265}{\ifmmode\geq\else{$\geq$}\fi}
\DeclareUnicodeCharacter{27E8}{\ifmmode\langle\else{$\langle$}\fi}
\DeclareUnicodeCharacter{2297}{\ifmmode\otimes\else{$\otimes$}\fi}
\DeclareUnicodeCharacter{27E9}{\ifmmode\rangle\else{$\rangle$}\fi}

\def\BibTeX{{\rm B\kern-.05em{\sc i\kern-.025em b}\kern-.08em
    T\kern-.1667em\lower.7ex\hbox{E}\kern-.125emX}}
\begin{document}

\title{Cyber Threat Detection Enabled by Quantum Computing}

\author{
\IEEEauthorblockN{Zisheng Chen}
\IEEEauthorblockA{\textit{Johns Hopkins University} \\
\textit{Information Security Institute}\\
Baltimore, USA \\
zchen215@jh.edu}
\and
\IEEEauthorblockN{Zirui Zhu}
\IEEEauthorblockA{\textit{Johns Hopkins University} \\
\textit{Information Security Institute}\\
Baltimore, USA \\
zzhu87@jh.edu}
\and
\IEEEauthorblockN{Xiangyang Li}
\IEEEauthorblockA{\textit{Johns Hopkins University} \\
\textit{Information Security Institute}\\
Baltimore, USA \\
xyli@jhu.edu}
}

\maketitle

\thispagestyle{plain}
\pagestyle{plain}

\begin{abstract}
Threat detection models in cybersecurity must keep up with shifting traffic, strict feature budgets, and noisy hardware, yet even strong classical systems still miss rare or borderline attacks when the data distribution drifts. Small, near-term quantum processors are now available, but existing work rarely shows whether quantum components can improve end-to-end detection under these unstable, resource constrained conditions rather than just adding complexity. We address this gap with a hybrid architecture that uses a compact multilayer perceptron to compress security data and then routes a few features to 2–4 qubit quantum heads implemented as quantum support vector machines and variational circuits. Under matched preprocessing and training budgets, we benchmark these hybrids against tuned classical baselines on two security tasks, network intrusion detection on NSL-KDD and spam filtering on Ling-Spam, and then deploy the best 4-qubit quantum SVM to an IBM Quantum device with noise-aware execution (readout mitigation and dynamical decoupling). Across both datasets, shallow quantum heads consistently match, and on difficult near-boundary cases modestly reduce, missed attacks and false alarms relative to classical models using the same features. Hardware results track simulator behavior closely enough that the remaining gap is dominated by device noise rather than model design. Taken together, the study shows that even on small, noisy chips, carefully engineered quantum components can already function as competitive, budget-aware elements in practical threat detection pipelines.
\end{abstract}


\section{Introduction}
Cyber threat detection rarely operates in a steady state. Traffic patterns drift, attack campaigns adapt, labels remain imbalanced, and yet intrusion detection systems (IDS) and spam filters are still expected to catch rare, borderline events under tight feature and compute budgets. Classical machine learning has greatly strengthened these systems, but even tuned neural and ensemble models can become brittle when benign and malicious behavior overlap in feature space or when models trained on one regime are pushed into another. In parallel, quantum machine learning (QML) models that offload part of the computation to small quantum chips promises richer decision boundaries under very compact encodings. What is missing is a clear understanding of whether such quantum components actually reduce errors in these unstable, resource-constrained regimes, or simply introduce new sources of complexity and noise. 

We therefore treat QML-enabled threat detection as an engineering gap rather than a theoretical curiosity and organize this work around two research questions: 

\begin{itemize}
    \item How to develop and train QML models for threat detection on high-dimensional network and text data when only a few noisy qubits are available?
    \item How do the resulting models perform on current quantum hardware compared with strong classical baselines and with idealized simulators?
\end{itemize}

Addressing these questions requires a pipeline-level design that keeps classical and quantum variants aligned from preprocessing and feature compression through the classical–quantum interface and circuit architecture to device-level noise handling, so that any observed gains or failures can be attributed to the quantum components rather than to incidental differences in training or evaluation. 

To make this comparison concrete, we build a hybrid architecture in which a compact multilayer perceptron (MLP) encoder compresses raw security data into a low dimensional representation that is then fed either to classical heads or to few qubit quantum heads: quantum support vector machines (QSVMs) based on a fidelity kernel and variational quantum classifiers (VQCs). We evaluate these hybrids on two representative tasks, network intrusion detection on NSL-KDD and spam classification on Ling-Spam, under matched preprocessing, feature budgets, and regularization, and we vary qubit count and circuit depth to expose the trade-offs between expressiveness and hardware noise. Finally, we deploy the best four qubit QSVM on an IBM Quantum backend with readout mitigation and dynamical decoupling, quantifying how much of the simulated behavior carries over once calibration drift, finite shots, and device constraints are taken into account. 

Our experiments indicate that shallow, noise-aware quantum classifiers can be competitive with tuned classical IDS baselines under strict feature and qubit budgets and, in some decision-boundary-sensitive regimes, can modestly reduce missed attacks and false alarms without increasing model size. At the same time, the gap between simulator and hardware is systematic but tractable when the encoder–circuit interface, transpilation, and noise controls are carefully engineered. This leads to three contributions: 
\begin{enumerate}
    \item An end-to-end empirical study of hybrid classical–quantum architectures for concrete cybersecurity tasks that explicitly avoids common pitfalls such as class imbalance and feature leakage
    \item A controlled comparison of QSVM and VQC heads against strong classical models that isolates when quantum components add value beyond a classical encoder
    \item A hardware-validated assessment of the current readiness of QML-enabled threat detection, highlighting the design choices, including feature compression, circuit structure, and noise mitigation, that matter most for near-term deployments
\end{enumerate}

\section{Literature Review}
QML is an emerging field at the intersection of quantum computing and machine learning. QML aim to leverage quantum phenomena such as superposition, entanglement, and interference to address challenges in data analysis, optimization, and pattern recognition. This section reviews the existing literature across various key dimensions

\subsection{Foundational algorithms}
The theoretical foundation of QML rests on algorithms that could potentially accelerate key computational tasks that are critical to classical machine learning. A landmark achievement in this area is the Harrow-Hassidim-Lloyd (HHL) algorithm, which offers an exponential speedup for solving large systems of linear equations, which is a common bottleneck in many classical machine learning methods for data analysis and prediction \cite{Biamonte2017QML}. While HHL's practical implementation requires fully error-corrected quantum computers due to its substantial hardware demands, it demonstrated that quantum computing could theoretically revolutionize computational workflows in machine learning.

For today's quantum hardware, known as the Noisy Intermediate-Scale Quantum (NISQ) era, the most relevant concepts are \textbf{quantum feature maps} and \textbf{quantum kernel methods}. These approaches do not aim for outright algorithmic speedup, but for the ability to create exceptionally rich and complex data representations. As Havlíček and colleagues showed in their pioneering experiment, we can use quantum circuits to transform classical data points into quantum states, effectively mapping them into a high-dimensional feature space native to quantum systems \cite{Havlicek2019QuantumFeatureSpaces}. The "similarity" between two data points in this quantum feature space, measured by a quantum kernel, can capture intricate patterns that might be very difficult for classical computers to detect.

This conceptual framework directly enables two practical QML approaches:
\begin{itemize}
    \item \textbf{Quantum Support Vector Machines (QSVMs)} that use the quantum kernel for classification.
    \item \textbf{Variational Quantum Classifiers (VQCs)} that combine the data mapping with a trainable quantum circuit.
\end{itemize}

Both approaches are naturally suited to classification problems found in cybersecurity, such as distinguishing malicious from benign network traffic or identifying spam. The hypothesis is that the quantum feature space might model the subtle and complex patterns of cyber threats more effectively. Our research, which tests both QSVM and VQC models on security datasets, provides a direct evaluation of these foundational algorithms in a practical context.

\subsection{Hybrid approaches and variational methods}
Given the limitations of current quantum hardware, including susceptibility to noise and limited qubit counts, the most practical path forward for QML relies on \textbf{hybrid quantum-classical architectures}. These approaches strategically combine the strengths of both classical and quantum computing to work around current hardware constraints. As Preskill articulated in his description of the NISQ era, these hybrid algorithms represent the most viable near-term strategy for extracting value from quantum processors, even before achieving a decisive quantum advantage \cite{Preskill2018QMLinNISQ}.

The foundation of these hybrid architectures is the \textbf{Variational Quantum Algorithm (VQA)}. In a VQA, a parameterized quantum circuit, also often called a quantum neural network or an ansatz, handles the quantum processing. The power of this circuit depends on the values of its adjustable parameters. A classical computer then optimizes these parameters, using feedback from the quantum device to minimize a cost function that defines the machine learning task.

However, this framework introduces unique challenges. The phenomenon of \textbf{barren plateaus}, where the training landscape becomes exponentially flat, making optimization practically impossible. This is a particular concern for randomly initialized, deep circuits \cite{McClean2018BarrenPlateaus}. Comprehensive reviews, such as the one by Cerezo et al., document these challenges and emphasize that careful circuit design is crucial for successful training on real devices \cite{Cerezo2023ChallengesOpportunities}.

Research into the fundamental capabilities of these models also shows a double-edged sword. Studies by Abbas et al. and Beer et al. confirm that certain quantum circuits possess high expressibility, meaning they can represent complex functions that would be challenging for classical models \cite{Abbas2021QuantumNeuralNetworks}\cite{Beer2020TrainingDeep}. However, this very expressibility often comes hand-in-hand with severe trainability issues.

Our project's design directly embraces this hybrid paradigm. We use a classical neural network (MLP) as a powerful feature extractor, reducing the dimensionality of the raw cybersecurity data. This processed data is then fed into a relatively shallow parameterized quantum circuit (VQC) or a quantum kernel method (QSVM) for the final classification. This split approach is a community-standard tactic: it uses established classical methods for heavy data lifting while reserving the quantum resource for what it might do best, which is creating sophisticated decision boundaries on well-prepared data. This balances the potential expressivity of quantum models with the practical necessity of keeping them trainable on today's hardware.

\subsection{Quantum data encoding strategies}
A critical and often underestimated step in any QML pipeline is data encoding, the process of transforming classical data into a quantum state. The choice of encoding strategy has major implications for both the model's expressive power and the model’s feasibility on near-term hardware. Different schemes create a direct tradeoffs between the richness of the data representation and the quantum resources required.
Common techniques include :

\begin{itemize}
    \item \textbf{Basis encoding}: representing classical bits directly as quantum basis states
    \item \textbf{Amplitude encoding}: storing data in the complex probability amplitudes of a quantum state
    \item \textbf{Angle encoding}: mapping features to rotation angles of individual qubits.
\end{itemize}

Each method has distinct advantages and drawbacks. Amplitude encoding is highly qubit-efficient but requires quantum circuits that are generally too deep for current processors. Angle encoding is less compact, which results in much shallower and more NISQ-friendly circuits.

This resource constraint has led to increased focus on \textbf{hybrid encoding strategies}. As explored in foundational texts, one approach is to first use classical preprocessing, such as principal component analysis (PCA) or autoencoders, to distill the most salient features from high-dimensional data \cite{Schuld2015Intro}. These reduced features can then be efficiently encoded into a small number of qubits using rotation gates, making the problem tractable for current devices \cite{Huang2021PowerOfData}.

The design of the quantum feature map itself is another crucial dimension. As demonstrated by Havlíček et al., carefully constructed feature maps can create kernel functions that are believed to be difficult to simulate classically \cite{Havlicek2019QuantumFeatureSpaces}. However, these powerful maps often require significant entanglement and circuit depth. A more recent innovation, \textbf{data re-uploading}, offers a way to increase model complexity without adding more qubits by repeatedly encoding the data throughout the quantum circuit, though this increases exposure to noise \cite{PerezSalinas2020DataReupload}.

For our cybersecurity application, the literature strongly validates our planned methodology. The high dimensionality of network traffic and text data makes pure quantum encoding impractical. Therefore, our strategy of employing classical preprocessing for dimensionality reduction, followed by a shallow, angle-based encoding into a small qubit register, represents a necessary and well-supported compromise to bridge the gap between complex data and NISQ-era limitations.

\subsection{Trainability and optimization issues}
The practical success of variational quantum algorithms heavily depends on our ability to effectively train them, but this process faces fundamental challenges that are different from classical deep learning. The most prominent of these challenges is the barren plateau phenomenon, where the cost function's gradients vanish exponentially as the number of qubits increases, effectively halting the optimization process \cite{McClean2018BarrenPlateaus}. This occurs becausse the parameter landscape becomes overwhelmingly flat for many randomly initialized quantum circuits, making it impossible to determine which direction leads toward improvement.

Research has identified several strategies to address this challenge. Cerezo et al. and others have shown that using local cost functions rather than global measurements, along with carefully structured problem-inspired ansatze, can help maintain measurable gradients \cite{Cerezo2023ChallengesOpportunities}. Similarly, Beer et al. demonstrated that strategic parameter initialization and constraining circuit depth are essential in maintaining trainability \cite{Beer2020TrainingDeep}. These findings directly inform our experimental design, where we limit both qubit count and circuit depth to avoid barren plateaus.

In addition to gradient issues, quantum models show substantial run-to-run variance due to the random initialization of parameters and the stochastic nature of quantum measurements. This variability motivates extensive hyperparameter tuning and multiple training runs with different random seeds to obtain statistically meaningful results \cite{Abbas2021QuantumNeuralNetworks}. This is a requirement we incorporated into our experimental protocol through multiple random restarts and robust statistical analysis.

The optimization process itself also presents unique difficulties. Classical optimizers must navigate a landscape characterized by numerous local minima and subtle correlations between parameters. This complexity highlights the value of strategies like the classical pretraining of feature extraction layers in a hybrid model, which can help to position the quantum circuit in a more favorable region of the parameter space before fine-tuning and improving the convergence of the hybrid quantum-classical model \cite{Cerezo2023ChallengesOpportunities}.

\subsection{Applications in cyber security}
The application of QML in cybersecurity is a developing but rapidly growing field of research. The potential of quantum computation to process complex, high-dimensional data has sparked interest in its use for detecting sophisticated cyber threats that often evade classical systems. Current literature in this domain is characterized by promising proof-of-concept studies, though significant gaps remain between theoretical potential and practical implementation. 

Several recent studies have begun exploring this intersection:

Gouveia and Correia (2019) provided an early investigation into quantum-assisted network intrusion detection using simulated environments, focusing primarily on unsupervised learning approaches \cite{Gouveia2019NetworkID}. Their work demonstrated the conceptual feasibility but highlighted the substantial challenges in translating these ideas to practical systems.

Kalinin et al. (2022) conducted a broader experimental survey that applied various QML approaches to intrusion detection contexts. This provided valuable insights into the practical constraints and performance boundaries of current quantum approaches \cite{Kalinin2022SecurityID}. Their findings emphasized the critical importance of data preprocessing and feature selection for quantum methods.

More recently, Abreu et al. (2024) proposed a hybrid quantum-classical intrusion detection system with comprehensive simulation experiments on public datasets \cite{Abreu2024QMLIDS}. While they reported competitive performance in noiseless simulation environments, their work notably acknowledged the limitations of current hardware validation.

Kim et al. (2024) developed a quantum approach using outlier analysis for intrusion detection, showing promising results on limited datasets but also highlighting the scalability challenges facing current quantum methods \cite{Kim2024OutlierAnalysis}.

Across these studies, several consistent patterns can be observed, which inform our research methodology. First, all successful implementations rely heavily on dimensionality reduction and sophisticated preprocessing to make problems tractable for systems with limited qubits. Second, the field remains dominated by simulation-based results, with very limited validation on actual quantum hardware. Third, there is a notable absence of rigorous analysis regarding the adversarial robustness of QML-based security systems.

These characteristics in the existing literature strongly validate our experimental approach. The selection of standardized datasets like NSL-KDD and Ling-Spam allows for direct comparison with both classical baselines and emerging quantum approaches. More importantly, our explicit focus on comparing simulator versus hardware results directly addresses the most significant gap in current research, moving beyond pure simulation to assess real-world feasibility in controlled, reproducible experiments.

\subsection{Noise, error mitigation, and hardware constraints}
The performance of QML models on current NISQ-era hardware is fundamentally constrained by various sources of noise and hardware limitations. Understanding these constraints is essential for realistic assessment of QML's practical potential in cybersecurity applications.

Quantum processors today face multiple challenges including gate infidelity, decoherence (characterized by $T_1$ and $T_2$ times), measurement errors, crosstalk, and calibration drift \cite{Preskill2018QMLinNISQ}. These noise sources limit the feasible circuit depth and qubit count for practical algorithms. As noted by Cerezo et al., the interplay between these noise types creates complex operational constraints that significantly impact model performance \cite{Cerezo2023ChallengesOpportunities}.

Since full quantum error correction remains impractical for current devices, error mitigation is still the primary strategy for obtaining meaningful results. These techniques operate on the principle of characterizing and correcting for specific error types without the overhead of full error correction. 

Key approaches include:
\begin{itemize}
    \item \textbf{Measurement Error Mitigation}: calibrating and correcting for readout errors through classical post-processing of measurement results
    \item \textbf{Zero-Noise Extrapolation (ZNE)}: intentionally scaling noise levels to extrapolate back to zero-noise expectations
    \item \textbf{Probabilistic Error Cancellation}: applying corrective operations based on characterized noise models
    \item \textbf{Dynamic Decoupling}: using pulse sequences to protect qubits from environmental noise during computation
\end{itemize}
While these methods do not provide provable correctness like error correction, they have demonstrated practical utility in improving result quality on current hardware \cite{Preskill2018QMLinNISQ}.

For our cybersecurity classification experiments, we account for these hardware realities through several design choices. We limit circuit depth and qubit count to match the capabilities of available IBM Quantum processors. We also compare performance between noiseless simulation and physical hardware to quantify the "reality gap" that current error mitigation techniques can cover.

This approach allows us to assess not just theoretical potential but practical feasibility, which provides crucial insights into how close QML is to real-world cybersecurity deployment given current hardware constraints and mitigation capabilities.

\subsection{Scalability and future directions}
The trajectory of QML points toward progressively more capable systems, although the path from current NISQ devices to practical advantage would take multiple evolutionary stages. Research into scalability addresses both the fundamental requirements for quantum advantage and the practical pathway for integrating quantum methods into classical workflows.

Theoretical work has begun establishing the conditions under which QML could achieve provable advantages over classical approaches. Under idealized, fault-tolerant assumptions, carefully constructed quantum models, particularly in kernel methods and specific learning architectures, could provide computational benefits for certain problem classes. Liu et al. (2021) demonstrated that rigorous quantum speedups are possible in supervised learning settings, but these advantages typically are more apparent in carefully constructed scenarios \cite{Liu2021Supervised}.

In the near to medium term, the field's development will likely follow a pattern of progressive hybrid migration. As hardware improves in scale, fidelity, and connectivity, we can expect increasingly sophisticated quantum components being integrated into classical machine learning systems. This evolution might begin with handing specific computational bottlenecks in classical algorithms before progressing to more comprehensive approaches native to quantum.

The ultimate scalability of QML for cybersecurity will depend on concurrent advances across multiple domains. Hardware improvements enable deeper circuits with more qubits, better error mitigation techniques can extend the useful depth of computations, and algorithmic innovations will be able to use quantum resources more efficiently. Our work contributes to this trajectory by establishing baseline performance metrics and identifying the most promising architectural patterns for future development.

\section{Problem Definition}
While quantum computing offers theoretical advantages for handling complex data patterns through superposition and entanglement, its practical application to cybersecurity tasks faces significant challenges in the NISQ era. Current quantum hardware limitations, including qubit noise, decoherence, and limited availability, restrict the implementation of complex quantum circuits for security applications. Also, substantial integration challenges become apparent when combining quantum components with classical security infrastructure and require careful design of hybrid architectures and data processing pipelines.

The central research challenge we address is not whether quantum models can outperform classical models in idealized settings, but whether QML can be realistically and meaningfully applied to cybersecurity tasks given current technological constraints. This investigation focuses on understanding under what specific system configurations QML approaches demonstrate practical utility for security applications. Our research is guided by the following two fundamental questions:

\begin{itemize}
    \item \textbf{How do we develop and train QML models of threat detection for cybersecurity?} This includes the practical challenges of designing quantum circuits for security tasks, selecting appropriate data encoding strategies, implementing effective training procedures despite quantum noise, and creating viable hybrid quantum-classical architectures that take advantage of the strengths of both paradigms.
    \item \textbf{How well do the trained QML models perform on current quantum computing hardware?} This addresses the critical performance gap between theoretical simulation and practical implementation, in addition to examining how factors like qubit count, noise mitigation strategies, and quantum hardware selection affect model accuracy, reliability, and scalability in real-world security scenarios.
\end{itemize}

By systematically investigating these questions, this research aims to provide concrete insights into the current readiness of QML technologies for cybersecurity applications.

\section{Basics of Quantum Computing}

Classical computation represents information with bits and transforms them with Boolean circuits. Quantum computation instead uses \emph{qubits} and \emph{unitary} transformations. This section serves to provide the essential background on quantum computing necessary to understand the QML approaches discussed in this work. The mathematics involved is fundamentally linear algebra, involving dealing with vectors, matrices, and inner products, but key physical rules like superposition, interference, and entanglement change what is representable and how information flows.

In this section, we build the story from states to circuits, then discuss noise and simulation, highlighting connections to machine learning. We recommend interested readers consult the textbook by Nielsen and Chuang for more detailed explanations

\subsection{From Classical Bits to Qubit States}
\textbf{Single qubit state:}
Analogous to the role of bits in classical computation, the basic element in quantum computation is the quantum bit(qubit)\cite{Cong2019QCNN}. A classical bit is either $0$ or $1$. A qubit is a unit vector in a two-dimensional complex Hilbert space,
\begin{gather*}
\ket{\psi} = \alpha\ket{0} + \beta\ket{1}\qquad \alpha,\beta\in\mathbb{C},\;|\alpha|^2+|\beta|^2=1 \\[6pt]
\ket{0} = \begin{bmatrix}1\\[2pt]0\end{bmatrix}\qquad\qquad\qquad \ket{1} = \begin{bmatrix}0\\[2pt]1\end{bmatrix}\quad 
\end{gather*}

A probabilistic bit uses probabilities $(p,1-p)$ over $\{0,1\}$. A qubit also yields probabilities upon
measurement, but it stores complex amplitudes whose relative phase governs interference
(a global phase is irrelevant). A single-qubit pure state admits the Bloch-sphere form
$\ket{\psi}=\cos(\tfrac{\theta}{2})\ket{0}+e^{i\phi}\sin(\tfrac{\theta}{2})\ket{1}$.

\textbf{Multi-qubit states:}
For $n$ qubits, the Hilbert space is the tensor product
$\mathcal{H}_n=(\mathbb{C}^2)^{\otimes n}\cong\mathbb{C}^{2^n}$ with the
computational basis
$\{\ket{z}: z\in\{0,1\}^n\}$, where
$\ket{z}=\ket{z_{n-1}}\otimes\cdots\otimes\ket{z_0}$.
A general pure state is a linear combination of these basis vectors:
\[
\ket{\Psi}=\sum_{z\in\{0,1\}^n}\alpha_z\,\ket{z},\qquad
\sum_{z}|\alpha_z|^2=1.
\]
By convention, amplitudes are ordered lexicographically by $z$.
\emph{Example ($n=2$).}
With basis $\{\ket{00},\ket{01},\ket{10},\ket{11}\}$,
\begin{gather*}
\ket{\Psi} = \sum_{b_1,b_0\in\{0,1\}}\alpha_{b_1b_0}\ket{b_1b_0} = \begin{bmatrix}
\alpha_{00}\\[2pt]\alpha_{01}\\[2pt]\alpha_{10}\\[2pt]\alpha_{11}
\end{bmatrix} \\[6pt]
\sum_{b_1,b_0}|\alpha_{b_1b_0}|^2=1.
\end{gather*}

\textbf{Density Matrices and Reduced States:}
Pure states correspond to $\rho=\ket{\Psi}\!\bra{\Psi}$, while mixed states are positive, trace-one operators.
For a bipartition $AB$, the state of subsystem $A$ is the partial trace $\rho_A=\mathrm{Tr}_B(\rho)$.
For pure bipartite states, $\rho_A$ is mixed if and only if $\ket{\Psi}$ is entangled (and likewise for $B$).
\textbf{ML Viewpoint:}
The vector $\ket{\Psi}\in\mathbb{C}^{2^n}$ is a high-dimensional embedding;
inner products $\braket{\Phi|\Psi}$ define similarities (kernels).
Product states correspond to feature factorizations; entangled states encode
high-order feature interactions without explicit feature engineering.

\subsection{Superposition and Interference}
A \textbf{superposition} is a linear combination of basis states. For a single qubit with orthonormal basis \(\{\ket{0}, \ket{1}\}\), a general state is:
\[
\ket{\psi} = \alpha \ket{0} + \beta \ket{1}, \qquad \alpha, \beta \in \mathbb{C}.
\]
The coefficients \(\alpha\) and \(\beta\) are probability amplitudes satisfying \(|\alpha|^2 + |\beta|^2 = 1\).

Measuring \(\ket{\psi}\) in \(\{\ket{0}, \ket{1}\}\) yields outcome 0 with probability 
\(|\alpha|^2\) and outcome 1 with probability \(|\beta|^2\). After measurement, the state collapses to \(\ket{0}\) (if outcome 0) or \(\ket{1}\) (if outcome 1).

Interference acts like learned constructive/destructive feature interaction at the amplitude level (not merely probabilities). Many quantum feature maps are Fourier-like, enabling expressive periodic decision boundaries with shallow circuits\cite{Cerezo2021CostBP}.

\subsection{Multi-Qubit Systems and Entanglement}
For $n$ qubits the state space is the tensor product $\left(\mathbb{C}^2\right)^{\otimes n}\cong\mathbb{C}^{2^n}$. 
A state is \emph{separable} (product) if it factorizes as
$\ket{\Psi}=\ket{\psi_{n-1}}\otimes\cdots\otimes\ket{\psi_0}$, otherwise it is
\emph{entangled}. For two qubits, reshape the amplitude vector into the
$2\times 2$ matrix

$
A=\begin{psmallmatrix}
\alpha_{00}&\alpha_{01}\\
\alpha_{10}&\alpha_{11}
\end{psmallmatrix}.
$

Then $\ket{\Psi}$ is separable iff $\mathrm{rank}(A)=1$, equivalently
$\alpha_{00}\alpha_{11}-\alpha_{01}\alpha_{10}=0$, otherwise it is entangled.
(For general bipartitions, the Schmidt decomposition plays the same role.)

Entanglement is not just strong correlation; it forbids any decomposition into local hidden states. It couples subsystems in a way classical probability cannot emulate without exponential resources. The Bell state:
\[
\ket{\Phi^+}=\tfrac{1}{\sqrt{2}}\big(\ket{00}+\ket{11}\big)
\]
has perfectly correlated outcomes, yet each qubit alone is maximally mixed (its reduced density matrix is $I/2$).

As a built-in mechanism, Entanglement couples features across subsystems, creating high-order interactions without explicit feature engineering,which means that entangling feature maps can implicitly lift data to an exponentially large space with short circuits, potentially separating classes that are hard for shallow classical models. In kernel and variational models alike, the right \emph{pattern} of entanglement
controls inductive bias: local patterns emphasize short-range correlations; global patterns can represent
long-range parity-like structure but are harder to train on noisy hardware\cite{Gorishniy2021Revisiting}.

Most QML circuits specify which qubits are entangled by a \emph{connectivity graph} $G=(V,E)$,
where $V=\{1,\dots,n\}$ and $E$ lists qubit pairs acted on by 2-qubit gates. Common choices include:

\begin{itemize}
  \item \textbf{Linear (nearest-neighbor chain)} $E=\{(1,2),(2,3),\dots,(n\!-\!1,n)\}$.
  Shallow, hardware-friendly, favors local feature mixing. Good default when input features have a known order
  (e.g., time, space, $n$-gram windows).

  \item \textbf{Circular (ring)} Linear plus $(n,1)$. Slightly stronger long-range coupling with minimal extra depth.

  \item \textbf{Full (all-to-all)} $E=\{(i,j): i<j\}$. Maximizes connectivity in one layer; expressive but depth/barren-plateau
  risks increase after transpilation to limited hardware.

  \item \textbf{Star (hub-and-spoke)} $E=\{(c,i): i\neq c\}$ for a central hub $c$. Efficiently aggregates many features into a
  \emph{global} summary at the hub; useful when one register acts as a ``classifier'' qubit.

  \item \textbf{Brickwork / ladder / 2D lattice} Alternating local pairs per layer (e.g., $(1,2),(3,4),\ldots)$ then shifted
  $(2,3),(4,5),\ldots$. Scales well, mirrors convolutional mixing with gradually increasing receptive field.

  \item \textbf{Hardware-native (e.g., heavy-hex, ion all-to-all)} Matches device couplings to reduce depth and noise;
  pick the densest pattern your device supports reliably.
\end{itemize}

Entanglement depth (size of the largest genuinely entangled subset) and entanglement entropy
help diagnose whether a circuit is mixing locally or globally. For noisy, shallow QML, modest depth with structured,
sparse entanglement often yields better trainability than fully global entanglement.

\subsection{Quantum Gates}
\label{subsec:q-gates}
Quantum \emph{gates} are unitary matrices ($U^\dagger U=I$), hence reversible linear maps on state vectors.
Common one-qubit gates include
\[
\begin{gathered}
X=\begin{bmatrix}0&1\\ 1&0\end{bmatrix},\quad
Z=\begin{bmatrix}1&0\\ 0&-1\end{bmatrix},\quad
H=\tfrac{1}{\sqrt{2}}\!\begin{bmatrix}1&1\\ 1&-1\end{bmatrix} \\[10pt]
R_z(\theta)=\begin{bmatrix}e^{-i\theta/2}&0\\ 0&e^{i\theta/2}\end{bmatrix}.
\end{gathered}
\]
Two-qubit gates (e.g., CNOT) create entanglement. Sets like $\{H,T,\mathrm{CNOT}\}$ are universal, i.e., can approximate any unitary.

A quantum gate is just a matrix multiply on vector(n-qubits state).
Local gates act as tensor products, e.g.\ $U_0\otimes U_1\otimes\cdots\otimes U_{n-1}$,
while two-qubit gates (CNOT, CZ) correlate subsystems and can \emph{create}
entanglement from product inputs.

If each parameterized gate is of the form
\[
U_j(\theta_j)=e^{-i\frac{\theta_j}{2} G_j},\qquad G_j^\dagger=G_j,
\]
then the model output $f_\theta(x)$ is a smooth function of $\theta$. For \emph{Pauli-generated} rotations
($G_j^2=I$), the parameter-shift rule gives an \emph{exact}, hardware-compatible gradient:
\[
\begin{gathered}
\frac{\partial f_\theta(x)}{\partial \theta_j}
= \tfrac{1}{2}\Big[f_{\theta^{(j,+)}}(x)-f_{\theta^{(j,-)}}(x)\Big]\\[6pt]
\theta^{(j,\pm)}=\theta\ \text{with}\ \theta_j\mapsto \theta_j\pm\tfrac{\pi}{2}.
\end{gathered}
\]

More generally, if $G_j$ has two eigenvalues $\pm r$, the shift becomes $\pm \tfrac{\pi}{2r}$ and the prefactor $r$:
\[
\frac{\partial f}{\partial \theta_j}
= \frac{r}{2}\Big[f_{\theta^{(j,+)}}-f_{\theta^{(j,-)}}\Big].
\]
Thus, to get $\nabla_\theta \mathcal{L}$ for a loss $\mathcal{L}(f_\theta(x),y)$, we evaluate the same circuit at two shifted values of $\theta_j$ per parameter (expectation estimates share the same measurement interface). This yields unbiased stochastic gradients when expectations are estimated with finite shots.

By analogy with traditional methods, we can consider quantum gates as linear (unitary) layers; tensor products provide channel-wise expansion; controlled/two-qubit gates are feature-interaction layers; the readout is a linear functional (observable) applied to the representation; and
parameter-shift delivers backprop-compatible gradients on real hardware\cite{Wierichs2022ParamShift}.

\subsection{Quantum Circuits}
A quantum circuit composes gates in time, optionally ending with measurement.\,
\begin{center}
\begin{quantikz}
\lstick{$\ket{0}$} & \gate{H} & \ctrl{1} & \meter{} \cw \rstick{$c_0$} \\
\lstick{$\ket{0}$} & \qw      & \targ{}  & \meter{} \cw \rstick{$c_1$}
\end{quantikz}
\end{center}
This circuit prepares $\ket{\Phi^+}$: $H$ creates superposition, CNOT entangles. In ML terms: (i) \emph{encoding} layer maps data $x\mapsto \ket{\phi(x)}$, (ii) \emph{variational} layers $U(\theta)$ mix/entangle features, (iii) \emph{readout} measures observables as predictions. Depth/width trade-offs mirror neural nets (capacity vs.\ optimization/noise sensitivity)\cite{Henderson2020Quanvolution} .

\subsection{Measurement and Shots}
\label{subsec:measurement-shots}

Quantum algorithms end by \emph{measuring} qubits to extract classical information.
For an $n$-qubit state $\ket{\psi}\in \mathbb{C}^{2^n}$, a projective measurement in the computational ($Z$) basis
uses the rank-1 projectors $\{\Pi_z=\ket{z}\!\bra{z}\}_{z\in\{0,1\}^n}$.
The \textbf{Born rule} gives the outcome distribution:
\[
\begin{gathered}
p(z)=\Pr[\text{outcome } z]=\braket{\psi|\Pi_z|\psi}=\big|\braket{z|\psi}\big|^2\\[6pt]
\sum_{z} p(z)=1.
\end{gathered}
\]
A general measurement (POVM) is a set $\{E_m\}_{m}$ of positive semidefinite operators with $\sum_m E_m=I$; the
probability of outcome $m$ is $\Pr[m]=\bra{\psi}E_m\ket{\psi}$ (projective measurements are a special case).
In practice, most NISQ algorithms use projective measurements in $Z$ after optional basis changes: to measure
$X$ on qubit $j$, apply $H$ to $j$ and measure $Z$; to measure $Y$, apply $S^\dagger H$ then measure $Z$.

\textbf{Shots:}
A measurement is destructive, where one run of the circuit yields a single bitstring $z\in\{0,1\}^n$. To estimate probabilities or expectations, we repeat the same circuit $S$ times and each repetition is a \textbf{shot}.
Let $\{z^{(s)}\}_{s=1}^S$ be the observed bitstrings and $\hat{p}(z)=\tfrac{1}{S}\sum_{s}\mathbf{1}\{z^{(s)}=z\}$ the empirical frequency.

\textbf{Estimating Expectations From Shots:}
Many models predict via the expectation of an observable $O$ (Hermitian, $\|O\|\le 1$), e.g.\ a Pauli or a sum of Paulis.
For a single-qubit Pauli $Z$ measured on qubit $j$, the outcome per shot is $o^{(s)}\in\{+1,-1\}$ and
\[
\begin{gathered}
\mu \;=\; \bra{\psi}Z_j\ket{\psi} \;=\; \mathbb{E}[o],\qquad
\hat{\mu} \;=\; \frac{1}{S}\sum_{s=1}^S o^{(s)}\\[6pt]
\mathrm{Var}[\hat{\mu}] \;=\; \frac{\mathrm{Var}[o]}{S}\le \frac{1}{S}.
\end{gathered}
\]
Thus the standard error scales as $\mathcal{O}(1/\sqrt{S})$ (classical Monte Carlo).
For a parity operator like $Z_i\!\otimes\!Z_j$, compute $o^{(s)}=(-1)^{z_i^{(s)}\oplus z_j^{(s)}}$ from each bitstring and average.
For a sum $O=\sum_\ell w_\ell P_\ell$ of commuting Paulis $\{P_\ell\}$ measured in the same basis,
estimate each $\hat{\mu}_\ell$ from the same shots and return $\widehat{\langle O\rangle}=\sum_\ell w_\ell \hat{\mu}_\ell$.

\textbf{Number of Shots Needed:}
Hoeffding’s inequality for $\pm 1$ outcomes gives, with probability at least $1-\delta$,
\[
|\hat{\mu}-\mu|\le \epsilon \quad \text{if}\quad
S \;\ge\; \frac{1}{2\epsilon^2}\ln\!\frac{2}{\delta}.
\]
Example: for $\epsilon=0.01$ and $\delta=0.05$, $S\approx 18{,}500$ shots.
Decomposing $O$ into fewer commuting groups reduces total shots by measurement grouping.

\textbf{Measurement and ML readout:}
In variational classifiers, the scalar logit is often $f_\theta(x)=\bra{\psi(x;\theta)} O \ket{\psi(x;\theta)}$.
We estimate $f_\theta(x)$ by sampling, then pass it into a classical loss (e.g., logistic).
For multi-class outputs, measure a \emph{set} of commuting observables per class or use several readout qubits; non-commuting
readouts require separate shot pools (higher cost).

\textbf{Basis Rotation and Non-Commuting Observables:}
To measure a Pauli $P\in\{X,Y,Z\}^{\otimes n}$, apply a local unitary $V$ that maps $P\mapsto Z^{\otimes n}$, then measure $Z$.
Non-commuting terms cannot share shots; partition $\{P_\ell\}$ into commuting groups and measure each group in its own rotated basis.

\textbf{Gradients With Shots:}
With the parameter-shift rule (Sec.~\ref{subsec:q-gates}), each partial derivative is the difference of two \emph{expectations} at shifted angles.
We estimate both expectations with shots; the gradient variance scales as $\mathcal{O}(1/S)$, so increasing shots reduces both prediction and gradient noise.
On simulators, adjoint differentiation provides exact (shot-free) gradients; on hardware, tune shots per parameter adaptively.

\subsection{Quantum Error and Noise}
Real hardware is noisy: decoherence (finite $T_1/T_2$), gate infidelities, and measurement errors. Single-qubit noise channels admit Kraus forms, e.g.\cite{Nation2021M3}
\[
\mathcal{E}_{\text{depol}}(\rho)=(1-p)\rho+\tfrac{p}{3}\big(X\rho X+Y\rho Y+Z\rho Z\big).
\]
\textbf{Error Mitigation Strategies:}
Fault-tolerant error correction encodes one logical qubit into many physical qubits (e.g., surface code) and actively corrects errors in real time. Error mitigation (e.g., zero-noise extrapolation, probabilistic error cancellation) reduces bias without full error-correcting codes.\cite{Cai2023QEM}.

\textbf{ML Relevance:} Noise limits circuit depth and can flatten gradients (\emph{barren plateaus}), but it can also act like regularization when carefully managed. Practical QML often balances expressivity and noise via shallow, problem-informed ansatzes, robust cost functions, and batching many \emph{shots} (repeated measurements) to reduce variance\cite{McClean2018BarrenPlateaus}\cite{Pesah2021NoBPQCNN}.

\subsection{Quantum Simulation}
{Digital simulation decomposes $e^{-iHt}$ into gates (Trotter--Suzuki, qubitization, QSP), while analog simulation engineers a device with Hamiltonian $H_{\mathrm{sim}}\approx H$. Classical backends simulate circuits exactly (state vector), approximately (tensor networks for low entanglement), or via stabilizer methods (for Clifford circuits).

\textbf{ML Relevance:} Quantum simulators enable three key functions: prototyping QML pipelines before hardware deployment, exploring model classes like time-evolution layers and variational simulation\cite{Abbas2023OptimizationIDS}, and evaluating quantum kernels $k(x,x')=|\braket{\phi(x)|\phi(x')}|^2$ for QSVM methods.

\subsection{Takeaway (mapping to ML)}
\begin{itemize}
  \item \textbf{State}: $\ket{\psi}$  representation/embedding.
  \item \textbf{Superposition and interference}: expressive linear combos with phase-sensitive interactions.
  \item \textbf{Entanglement}: high-order feature coupling without explicit feature engineering.
  \item \textbf{Gates} (unitaries, often parameterized): trainable layers.
  \item \textbf{Circuit} (encode $\to$ entangle/mix $\to$ read out): computation graph / model.
  \item \textbf{Error/noise}: constraints shaping architecture/optimization (regularization-like effects if handled well).
  \item \textbf{Simulation}: development and evaluation stack for QML models and kernels.
\end{itemize}

\section{Dataset and Preprocessing}

\subsection{NSL-KDD}
We use the standard \texttt{KDDTrain+.txt} for training/validation and \texttt{KDDTest+.txt} for testing. Each record has 41 features with three categorical fields (\texttt{protocol\_type}, \texttt{service}, \texttt{flag}) plus \texttt{label} and \texttt{difficulty} (the latter is dropped). Labels are binarized as $y=1$ for any non-\texttt{normal} attack and $y=0$ otherwise. A stratified split on the training file creates a validation set (default 20\%), and all transformers are fit strictly on the training split to avoid leakage\cite{Tavallaee2009NSLKDD}. 

For preprocessing, numeric features are standardized; categorical features are frequency-bucketed (rare levels $\rightarrow$ \texttt{\_\_RARE\_\_}, min count 50) and then one-hot encoded with unknowns ignored. The resulting dense matrix is stored in \texttt{float32}. Class imbalance is handled via \texttt{class\_weight='balanced'} to derive per-class weights for the cross-entropy loss. 

\subsection{Ling-Spam}
Raw emails are loaded from the \texttt{lingspam\_public} tree; files whose names start with \texttt{spmsg} are labeled \texttt{spam}, others \texttt{ham}. After a stratified split into train/val/test (default test 20\% and validation 20\% of the remainder, no leakage), texts are vectorized using TF--IDF (lowercased, English stopwords, $\text{min\_df}=2$, $\text{max\_df}=0.95$, up to $n$-grams per configuration). TF--IDF outputs dense \texttt{float32} arrays. As with NSL--KDD, per-class weights are computed with \texttt{class\_weight='balanced'} \cite{Androutsopoulos2000LingSpam}. 

\section{Method}
With a highly constrained quantum budget (2–4 qubits), performance is best supported by (i) maintaining shallow circuit depth, (ii) temporally reusing qubits via data re-uploading, and (iii) assigning the majority of representation learning to a compact classical component.\cite{Huang2021PowerOfData} Based on these ideas, we design two models: MLP-QSVM and MLP-VQC.
\begin{enumerate}
    \item \textbf{Quantum-kernel methods (QSVM)}: A depth-limited feature map introduces nonclassical phases and entanglement to embed inputs into a higher-dimensional Hilbert space. Because optimization is performed entirely by the classical SVM, the approach mitigates barren-plateau phenomena and gradient shot-noise issues. 
    \item \textbf{Compact hybrid VQC quantum head}: Data re-uploading increases expressivity without increasing the qubit count, while shallow entangling layers capture correlations under strict depth constraints. A small classical head (or tail) stabilizes training and improves robustness to hardware noise.
\end{enumerate}

\subsection{Classical Component Model Architecture}
\label{sec:classical-part}
Our classical module is a compact multilayer perceptron (MLP) that produces both a normalized embedding and 2-class logits. Let $x\!\in\!\mathbb{R}^{d_{\text{in}}}$ denote the input feature vector and $\{h_1,\dots,h_L\}$ the hidden sizes.

We choose this MLP as the classical module because it balances capacity, stability, and compatibility with the downstream quantum stage. Each design choice supports a specific goal:

\begin{itemize}

\item \textbf{Capacity for the data regime:} A few Linear--LayerNorm--GELU--Dropout blocks provide sufficient nonlinearity for binary text discrimination without the sample complexity and overfitting risk of CNN, RNN, or Transformer architectures. The parameter count scales linearly with hidden width, allowing capacity to be tuned tightly to dataset size.

\item \textbf{Geometry for the quantum head:} The MLP produces an L2-normalized embedding \( e \in \mathbb{R}^{d_e} \) on the unit sphere. This yields three benefits: (i) bounded coordinates map cleanly to rotation angles in the data-encoding circuit, (ii) the first \( q \le 4 \) coordinates can be used directly as qubit inputs, and (iii) the spherical constraint results in a well-conditioned feature space that improves kernel centering and margin geometry.

\item \textbf{Resource efficiency:} Unlike convolutional or attention layers, fully connected blocks have predictable compute and memory footprints and low inference latency. This keeps the classical part lightweight so the computational bottleneck remains in quantum kernel construction, not in feature extraction.

\item \textbf{Regularization and training stability:} LayerNorm and GELU stabilize activations; Dropout and weight decay promote robust generalization. The architecture exhibits simple, well-behaved gradients---with no recurrence or attention softmax---which reduces tuning burden and variance across runs.

\item \textbf{Modular dual-head design:} Returning both logits and embedding \( e \) lets the same encoder support two consumers: a linear classifier for a pure classical baseline and the QSVM for the hybrid variant. This isolates quantum improvements to the decision layer while holding the feature extractor fixed.

\item \textbf{Interpretability and diagnostics:} The low-dimensional embedding enables clear error analysis---such as per-class clustering and nearest-neighbor inspection---and straightforward ablations (varying \( d_e \), width, depth) to understand where gains originate, without entangling those effects with sequence modeling choices.
    
\end{itemize}

In short, the MLP provides just-enough nonlinearity and a \emph{geometrically constrained} embedding that couples cleanly to a small-qubit quantum kernel, yielding a controllable, stable, and computationally efficient classical front end for our hybrid design.

\subsubsection{Backbone}
The backbone stacks $L$ identical blocks, each
\[
\begin{gathered}
    z_\ell \;=\; \mathrm{Dropout}_p\!\left(\mathrm{GELU}\!\big(\mathrm{LN}(W_\ell a_{\ell-1}+b_\ell)\big)\right)\\[6pt]
    \quad a_0=x,\; a_\ell=z_\ell,\;\ell=1..L,
\end{gathered}
\]
with $W_\ell\!\in\!\mathbb{R}^{h_\ell\times h_{\ell-1}}$ (and $h_0=d_{\text{in}}$), $b_\ell\!\in\!\mathbb{R}^{h_\ell}$, LayerNorm applied channel-wise, GELU nonlinearity, and dropout rate $p$.

\textbf{Linear Transformation (\(W_\ell a_{\ell-1}+b_\ell\)):}
The dense affine projection mixes all input features and changes the dimensionality from \( h_{\ell-1} \) to \( h_\ell \). This creates new directions in feature space where subsequent nonlinearities can operate to separate classes, making this the primary capacity-carrying step in each block.

\textbf{Layer Normalization (LN):}
Applied to a single sample, LN normalizes each hidden vector across its feature dimension:
\[
\mathrm{LN}(u) = \gamma \odot \frac{u - \mu(u)}{\sqrt{\sigma^2(u)+\varepsilon}} + \beta,
\]
where \( (\gamma,\beta) \) are learnable scale and shift parameters. LN maintains activation scales independently of batch size, which stabilizes deep networks and enables consistent behavior even with small or variable batches. Positioning LN before the nonlinearity ensures the activation function receives inputs with zero mean and unit variance, yielding predictable layer-to-layer responses.

\textbf{GELU Nonlinearity:}
The Gaussian Error Linear Unit gates inputs smoothly via the approximation:
\[
\begin{gathered}
    \mathrm{GELU}(t) = t\,\Phi(t)\\[6pt]
\approx 0.5\,t\Big(1+\tanh\!\big[\sqrt{2/\pi}\,(t+0.044715\,t^3)\big]\Big)
\end{gathered}
\]
with the exact form given by \( t\,\Phi(t) \), where \( \Phi \) is the standard normal CDF. Compared to ReLU, GELU retains small negative values instead of hard-zeroing them, producing smoother gradients, fewer “dead” units, and better compatibility with LN-normalized inputs.

\textbf{Dropout\(_p\):}
This operation applies an elementwise Bernoulli mask \( m \sim \mathrm{Bernoulli}(1-p) \) to the post-activation values:
\[
\mathrm{Dropout}_p(z) = \frac{m \odot z}{1-p}.
\]
It discourages co-adaptation of features and acts as a form of stochastic model averaging, thereby improving generalization. Placing dropout \emph{after} the activation (and after LN) preserves the calibrated scale established by LN and prevents LN from inadvertently “undoing” the masking effect.

\textbf{Interface To Output Heads:}
The final hidden state \( a_L \) feeds two downstream heads: (i) an embedding head that applies L2-normalization to produce a unit-sphere representation (ensuring geometry-controlled features) and (ii) a linear classification head that outputs 2-class logits. This separation design allows the same backbone to support both a pure classical classifier and the QSVM kernel interface.

\textbf{Design Rationale for the Layer Ordering (Linear $\rightarrow$ LN $\rightarrow$ GELU $\rightarrow$ Dropout):}
\begin{itemize}
    \item Linear projects/mixes features
    \item LN standardizes the pre-activation so the nonlinearity operates in a stable regime
    \item GELU injects smooth nonlinearity for expressiveness
    \item Dropout regularizes the resulting features without being re-normalized away
\end{itemize}
Stacking such blocks increases the effective function order while keeping activation scales well-behaved across depth.

\subsubsection{Embedding Head}
The final hidden activation $a_L$ is projected to an embedding
\[
\tilde{e}=W_e a_L+b_e\in\mathbb{R}^{d_e},\qquad
e=\frac{\tilde{e}}{\lVert \tilde{e}\rVert_2},
\]
with $W_e\!\in\!\mathbb{R}^{d_e\times h_L}$, $b_e\!\in\!\mathbb{R}^{d_e}$. The L2 normalization constrains the embedding to the unit hypersphere, which stabilizes downstream metric geometry. \cite{TabTransformer2020}
\(W_e\) and \(b_e\) are learned end-to-end as part of the network’s parameters by backpropagation from the task loss used by the classifier head. 
Thus \(W_e\) learns a task-aligned subspace and \(b_e\) recenters it; weight decay can be applied to \(W_e\) while the normalization removes scale degeneracy.

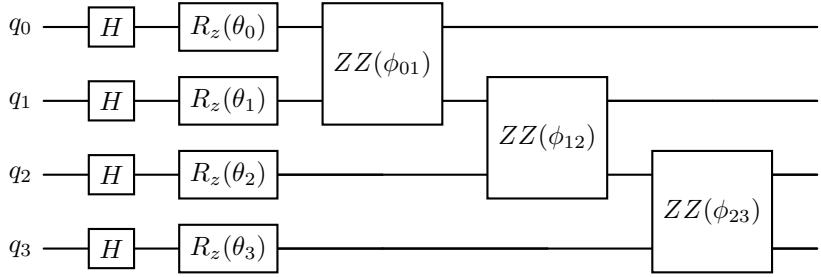
\begin{figure*}[t!]
\centering
\begin{quantikz}[row sep=0.35cm, column sep=0.6cm]
\centering
\lstick{$q_0$} & \gate{H} & \gate{R_z(\theta_0)} & \gate[wires=2]{ZZ(\phi_{01})} & \qw                           & \qw                           & \qw \\
\lstick{$q_1$} & \gate{H} & \gate{R_z(\theta_1)} & \ghost{ZZ(\phi_{01})}         & \gate[wires=2]{ZZ(\phi_{12})} & \qw                           & \qw \\
\lstick{$q_2$} & \gate{H} & \gate{R_z(\theta_2)} & \qw                           & \ghost{ZZ(\phi_{12})}         & \gate[wires=2]{ZZ(\phi_{23})} & \qw \\
\lstick{$q_3$} & \gate{H} & \gate{R_z(\theta_3)} & \qw                           & \qw                           & \ghost{ZZ(\phi_{23})}         & \qw
\end{quantikz}
\caption{ZZFeatureMap under 4 qubits}
\end{figure*}

\textbf{Design Rationale for Embedding Head}
The head design follows three principles. First, capacity and interface are decoupled: a separate projection \(W_e\) controls the embedding dimension \(d_e\) (and thus the model's bottleneck) independently of the backbone width \(h_L\). 

Second, spherical geometry is enforced via L2 normalization, placing \(e\) on the unit sphere and turning Euclidean distance into an angular metric; this stabilizes scales across layers and prevents any single coordinate from dominating downstream decisions. 

Third, a linear readout using classifier parameters \(W_c, b_c\) provides a pure classical baseline that interprets the spherical embedding via a margin model, while the same embedding \(e\) is reused by the quantum head. This ensures that improvements in feature quality benefit both classical and quantum variants without entangling their optimization objectives.

\textbf{Effect on the Quantum Feature Map}
The unit-sphere constraint gives a
\emph{bounded, well-conditioned interface}: each $e_i\in[-1,1]$ maps cleanly to
$\theta_i=\pi e_i$, avoiding out-of-range angles and keeping the feature map in a stable,
low-variance regime. Because $d_e$ is explicit, we can choose $d_e\!\ge\!q$ to feed exactly
$q$ qubits without additional adapters and treat $d_e$ as a controllable bottleneck that
suppresses spurious high-variance directions before kernelization. This combination improves
kernel centering and margin geometry for the quantum part while keeping the classical
and quantum consumers of $e$ perfectly aligned.

\subsubsection{Classification Head}
A lightweight linear head maps the embedding to logits
\[
\text{logits} = W_c\, e + b_c \in \mathbb{R}^{2},
\]
with $W_c\!\in\!\mathbb{R}^{2\times d_e}$, $b_c\!\in\!\mathbb{R}^{2}$. \(W_c\) and \(b_c\) are learned jointly with the encoder by backpropagation from the classification loss. The forward pass returns both $(\text{logits}, e)$ for reuse by quantum components. 

\subsubsection{Shape Summary}
$x\!\in\!\mathbb{R}^{d_{\text{in}}}\!\xrightarrow{\text{blocks}}\!a_L\!\in\!\mathbb{R}^{h_L}\!\xrightarrow{W_e}\!e\!\in\!\mathbb{R}^{d_e}$ (unit-norm)
$\xrightarrow{W_c}\!\mathbb{R}^{2}$. Hyperparameters exposed by the implementation are $(d_{\text{in}}, \{h_\ell\}_{\ell=1}^L, d_e, p)$.

\textbf{Design Rationale for Shape}
The linear readout is chosen for three aligned reasons. First, it provides minimal extra capacity with maximal comparability, adding the fewest possible parameters on top of the embedding ensures that any improvements can be attributed to the encoder or quantum head rather than to a heavy classifier, while also serving as a clean classical baseline that shares exactly the same embedding \(e\).

Second, it yields an angle-based decision geometry with \(\|e\|_2=1\), the margin for the positive class reduces to \(m = (w_1-w_0)^\top e + (b_1-b_0)\), so classification depends primarily on the angle between \(e\) and the vector \(v=w_1-w_0\). This naturally matches the geometry of kernel methods and simplifies analysis on the unit sphere.

Third, it ensures stable gradients and calibration, where the softmax–cross-entropy loss atop a single linear layer gives well-behaved gradients and avoids compounding nonlinearities after the normalization step.

\subsection{QSVM Model Architecture}
\label{sec:qsvm}
The quantum stage wraps a classical SVM around a \emph{precomputed quantum kernel} defined by a data-encoding circuit acting on a subset of the embedding coordinates.\cite{Schuld2014QSVM}\cite{Rebentrost2014QSVM}

\subsubsection{Data Encoding}
Given the MLP embedding $e\!\in\!\mathbb{R}^{d_e}$, Denote the truncated vector by $u\!\in\!\mathbb{R}^{q}$. Each feature $u_i$ is mapped to a rotation angle $\theta_i=\pi u_i$,\cite{Havlicek2019QuantumFeatureSpaces} and the angles parametrize a depth-$r$ entangling \textsc{ZZFeatureMap} on $q$ qubits:
\[
 \ket{\phi(u)}=U_{\text{ZZ}}(\theta; q, r, \textit{full-ent.})\ket{0}^{\otimes q}.
\]

The circuit alternates single-qubit $Z$-phase encodings with all-to-all $\mathrm{ZZ}$ entanglers; $r$ controls the number of repeated encoding–entangling layers and \textit{full entanglement} couples every qubit pair in each layer.

\textbf{ZZFeatureMap} The ZZFeatureMap offers several advantages: First, it captures explicit pairwise interactions via \(Z_i Z_j\) terms, moving beyond separable kernels from single-qubit rotations.

Second, it is stable and hardware-efficient, since all \(Z\)/\(ZZ\) operations commute, enabling low-depth scheduling, and the bounded angles (from unit-norm embeddings) keep gate fidelities well-conditioned.

Third, it provides a suitable inductive bias—while the classical MLP already mixes coordinates, the ZZ map emphasizes second-order relations among those mixed coordinates, offering useful nonlinearity for \(q \le 4\) qubits without overfitting.

Lastly, its capacity is controllable via repetition count \(r\), which expands the Fourier spectrum of the induced kernel (through more Pauli words) while keeping all parameters data-driven.

\textbf{Full Entanglement} Full entanglement provides several benefits: First, it imposes no artificial locality—our coordinates have no natural spatial order, so chain or ring entanglement would artificially privilege “neighbors.” Full entanglement couples every pair \((i,j)\) within each layer, matching the dense nature of the learned embedding.

Second, it optimizes information per qubit. With a small qubit budget (\(q\!\le\!4\)), the number of pairs \(q(q-1)/2\) is modest (at most 6), allowing us to afford all-to-all couplings and extract maximal pairwise structure per layer.

Third, it improves kernel alignment. All-to-all \(ZZ\) terms induce cross terms in the fidelity \(k(\theta,\theta')=|\langle\phi(\theta)|\phi(\theta')\rangle|^2\) that depend on angles between every pair of coordinates, enhancing margin geometry after kernel centering.

Finally, it balances expressivity against depth. Full entanglement at small \(q\) and shallow repetition count \(r\) yields richer features than sparse entanglement, with negligible extra depth overhead in simulation (and acceptable depth on hardware if \(r\) is kept small).

\subsubsection{Quantum kernel}
For samples $u$ and $v$, the kernel is the squared fidelity
\[
k(u,v)=\bigl| \langle \phi(u) \mid \phi(v) \rangle \bigr|^2,
\]
realized by preparing statevectors for $\ket{\phi(u)}$ and $\ket{\phi(v)}$ and taking the squared magnitude of their inner product. Gram matrices $K_{AA}$ and $K_{BA}$ are built for train and (val/test vs. train) sets, respectively.

\textbf{Centered Kernel Geometry}
Let $K\in\mathbb{R}^{n\times n}$ be the train–train Gram matrix and define the centering matrix $H=I-\tfrac{1}{n}\mathbf{1}\mathbf{1}^\top$.
We use the double-centering transform
\[
K_c \;=\; H K H,
\]
which subtracts the Hilbert-space mean feature vector so the SVM operates around the origin. For an out-of-sample block $K_{*A}\in\mathbb{R}^{m\times n}$ (validation/test vs.\ train), the consistent centering is
\[
\begin{aligned}
\tilde K_{*A} \;=\; K_{*A} &- \tfrac{1}{n}\,\mathbf{1}_m \mathbf{1}_n^\top K \\
                 &- \tfrac{1}{n}\, K_{*A} \mathbf{1}_n \mathbf{1}_n^\top \\
                 &+ \tfrac{1}{n^2}\,\mathbf{1}_m \mathbf{1}_n^\top K\, \mathbf{1}_n \mathbf{1}_n^\top .
\end{aligned}
\]

Because $k(\theta,\theta)=1$, no diagonal normalization is required before centering; after centering, the diagonal generally drops below $1$ (as in classical kernel PCA).\cite{Thanasilp2024ExponentialConcentration}

\begin{figure*}
    \centering
    \includegraphics[width=1\linewidth]{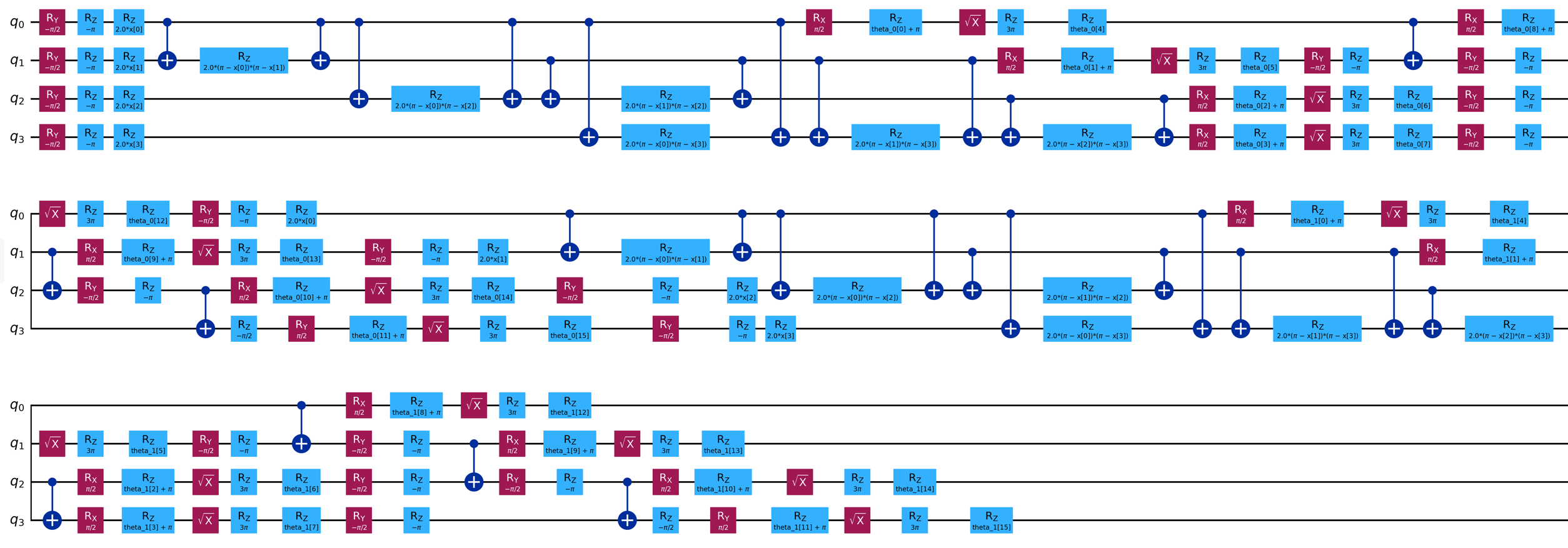}
    \caption{VQC Circuit Structure}
    \label{fig:vqc}
\end{figure*}

\textbf{Spectral/Expressivity View}
With a ZZ-type map, $U(\theta)$ is a product of commuting $Z$ and $ZZ$ phase operators.
Consequently, $k(\theta,\theta')$ expands into trigonometric polynomials of \emph{linear and pairwise} combinations of the angles
$\{\theta_i\}$ and $\{\theta'_i\}$.
Depth $r$ increases the highest accessible Fourier frequency and the number of Pauli words, widening the spectrum of features;
full entanglement ensures cross terms for \emph{every} pair $(i,j)$ appear in the kernel. This yields a controllable but rich hypothesis class without introducing trainable quantum parameters.

\textbf{Numerical/Structural Properties}
\begin{itemize}
\item \emph{PSD and symmetrization.} In exact arithmetic $K$ is PSD and symmetric. In practice, we enforce symmetry by $(K{+}K^\top)/2$
and, if needed, project to the PSD cone by clamping tiny negative eigenvalues to $0$ (finite-precision safeguard).
\item \emph{Bounds and scaling.} Each entry satisfies $0\le K_{ij}\le 1$.
The fidelity form makes the kernel automatically normalized ($K_{ii}=1$) and invariant to global phases.
\item \emph{Complexity.} Building $K$ for $n$ samples costs $O(n^2\,C_U)$ evaluations, where $C_U$ is the cost of one feature-map application
(statevector: $O(r\,2^q)$; hardware: circuit depth dominated by $r$ entangling layers). Memory is $O(n^2)$.
\item \emph{Shot estimation (hardware).} If estimating $k$ by sampling, the variance of a single estimate is $k(1-k)/S$ with $S$ shots;
achieving absolute error $\varepsilon$ uniformly over pairs requires $S=\Theta(\varepsilon^{-2})$, typically modest for small $q$.
\end{itemize}

\textbf{Interface to Our Embedding}
Because the classical embedding $e$ is unit-norm and we map coordinates to $Z$-phase angles, inputs to $U(\theta)$ are \emph{bounded} and well-conditioned.
This avoids extreme rotations and keeps the fidelity kernel numerically stable while preserving angular structure that the centered kernel can separate with a large margin.

\textbf{Summary}
The quantum kernel used here is the fidelity between states prepared by a ZZ feature map with full entanglement. It is PSD, bounded, and
explicitly encodes pairwise interactions among the $q$ selected coordinates; centering removes the mean feature and improves SVM geometry.\cite{Schuld2019FeatureHilbert}
Depth $r$ and qubit count $q$ control expressivity and cost, giving a compact yet expressive kernel for the hybrid classifier.

\subsubsection{Classical Margin Model}
A binary SVM with \texttt{kernel='precomputed'} and class balancing operates on the centered quantum Gram blocks. Its decision function $f(\cdot)$ is then applied to $\tilde{K}_{*A}$; no quantum parameters are learned and capacity is controlled by $(q, r)$ and the SVM margin.

\subsection{VQC Model Architecture}
\label{subsec:vqc}
In contrast to the fixed, kernel-defined geometry used by QSVM, the VQC replaces the non-parametric kernel with a \emph{trainable} quantum feature extractor that is co-optimized end-to-end with the classical front-end.\cite{Cerezo2021VQA}
The intent is twofold. First, it preserves the same data pathway, split protocol, and reporting metrics as in QSVM to enable controlled, apples-to-apples comparisons. Second, it introduces \emph{data re-uploading} and a lightweight, hardware-friendly entangling pattern so that expressive power can be increased under strict NISQ budgets on qubits and depth.\cite{Peruzzo2014VQE}

All design choices below are implemented in the accompanying program \texttt{hybrid\_cnn\_vqc\_nslkdd\_reupload.py} and exposed via command-line flags for reproducibility and operation.

\subsubsection{Data Encoding}
\label{subsec:vqc-encoding}
The encoding task is to map a preprocessed NSL-KDD sample $x\in\mathbb{R}^{d}$ to a length-$n_q$ vector of physical rotation angles that respects device constraints while preserving task-relevant signal. The encoder mirrors the QSVM front-end so that VQC and QSVM ingest identical tensors and differ only in the quantum back-end.

\textbf{Data Re-uploading}
To increase expressivity without growing entangling depth, the same angle vector $\alpha(x)$ is \emph{re-uploaded} $L$ times (flag \texttt{--vqc-reps}), i.e.,
\[
U(x,\theta)\;=\;\prod_{\ell=1}^{L}\, U_{\phi}\!\big(\alpha(x)\big)\;U_{\theta}^{(\ell)}.
\]
This preserves the embedding geometry while allowing trainable layers to refine the decision boundary.\cite{PerezSalinas2020DataReupload}

\textbf{Batching and Numerical Stability}
Angles are materialized as contiguous tensors, clipped to $[-\pi,\pi]$ post-quantization if shot noise is enabled, and optionally cached to avoid redundant host$\leftrightarrow$device transfers. Seeds (\texttt{--seed}) synchronize encoder initialization with the quantum layer to reduce variance across runs.

\subsubsection{Circuit structure}
\label{subsec:vqc-circuit}
To keep circuits shallow, hardware-friendly, and compiler-robust while enabling non-trivial entanglement, we combine a first-order entangling feature map with a minimal Two-Local ansatz and linear connectivity, matching the QSVM transpilation strategy to avoid target mismatch errors on common superconducting backends.

\textbf{Register and observable}
We allocate $n_q$ data qubits prepared in $\ket{0}^{\otimes n_q}$ and measure a tensor-product Pauli observable,
\[
\begin{gathered}
    \hat{O}\;=\;\bigotimes_{i=1}^{n_q} Z_i,\\[6pt]
    z(x;\theta)\;=\;\bra{0^{\otimes n_q}} U^\dagger(x,\theta)\,\hat{O}\,U(x,\theta)\ket{0^{\otimes n_q}}\in[-1,1],
\end{gathered}
\]
which serves as the scalar logit (fed into \texttt{BCEWithLogitsLoss}). A linear post-scale/bias may be learned jointly with the classical head.

\medskip
\textbf{Feature map $U_{\phi}$.}
We use a depth-1 entangling map that (i) applies $R_Z$/$R_Y$ data rotations from $\alpha(x)$ and (ii) inserts nearest-neighbor ZZ-style couplings (``\texttt{reps}=1''). This choice preserves locality and compiles well into modern native gate sets while providing pairwise correlations that QSVM kernels exploit implicitly.

\medskip
\textbf{Trainable ansatz $U_{\theta}^{(\ell)}$.}
Each variational block is a Two-Local layer with single-qubit rotations $\{R_y,R_z\}$ followed by linear entanglement using \texttt{CZ} (or its transpiled equivalent). With \texttt{--vqc-reps}$=L$, the full circuit is
\[
U(x,\theta) \;=\; \big[U_{\phi}(\alpha(x))\,U_{\theta}^{(1)}\big]\cdots\big[U_{\phi}(\alpha(x))\,U_{\theta}^{(L)}\big].
\]
\emph{Parameter count.} For $n_q$ qubits and one Two-Local layer per re-upload, the trainable parameter count is
\[
|\theta| \;=\; L \times \big(2n_q \;+\; (n_q-1)\times p_{\mathrm{ent}}\big),
\]
where $p_{\mathrm{ent}}$ is the number of entangling parameters per edge (zero for plain CZ). This linear scaling in $L$ makes re-uploading the main expressivity knob under a fixed depth budget.

\medskip
\textbf{Transpilation and target compatibility.}
Prior to estimator wrapping, circuits are decomposed to a conservative native set (e.g., \{\texttt{x, sx, rz, ry, rz, cx, cz, rzz, measure}\}) and transpiled with linear layout to reduce SWAP overhead. This mirrors the QSVM pipeline and avoids unsupported instructions; backend-specific passes (e.g., direction fixes) are enabled when available.

\medskip
\textbf{Depth/width budgeting and connectivity.}
We use \emph{linear} entanglement to align with typical device coupling maps and to keep two-qubit counts predictable:
\[
\begin{gathered}
    \text{2Q gates per block} \;\approx\; (n_q-1),\\[6pt]
    \text{depth} \;\approx\; L\cdot\big(c_1 + c_2(n_q-1)\big),
\end{gathered}
\]
for small constants $c_1,c_2$ set by the compiler. This matches the QSVM constraint envelope and enables fair runtime and error-rate comparisons.

\medskip

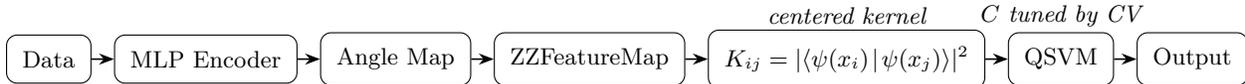
\begin{figure*}[t]
\centering
\begin{tikzpicture}[x=1cm,y=1cm]
  \node[node] (data) {Data};

  \node[node,right=0.3cm of data] (mlp) {MLP Encoder};
  \draw[arr] (data) -- (mlp);

  \node[node,right=0.3cm of mlp,align=center] (angles)
    {Angle Map};
  \draw[arr] (mlp) -- (angles);

  \node[node,right=0.3cm of angles,align=center] (zz)
    {ZZFeatureMap};
  \draw[arr] (angles) -- (zz);

  \node[node,right=0.3cm of zz,align=center] (K)
    {$K_{ij}=|\langle\psi(x_i)\!\mid\!\psi(x_j)\rangle|^2$};
  \node[op,above=2pt of K] {centered kernel};
  \draw[arr] (zz) -- (K);

  \node[node,right=0.3cm of K] (qsvm) {QSVM};
  \node[op,above=2pt of qsvm] {$C$ tuned by CV};
  \draw[arr] (K) -- (qsvm);

  \node[node,right=0.3cm of qsvm] (out) {Output};
  \draw[arr] (qsvm) -- (out);
\end{tikzpicture}
\caption{ Hybrid QSVM pipeline: Data $\rightarrow$MLP  $\rightarrow$ quantum kernel $\rightarrow$ QSVM.}
\end{figure*}

\textbf{Design Rational.}
The combination of (i) first-order entangling embeddings, (ii) shallow Two-Local blocks, and (iii) data re-uploading yields a controllable trade-off between expressivity and noise sensitivity, under precisely the same preprocessing, splits, and evaluation settings used for QSVM. Consequently, any performance deltas can be attributed to \emph{learned quantum features} rather than to differences in data handling or compilation.

\subsubsection{Learning objective and optimization.}
Let $y\in\{0,1\}$ be the binary label (\textit{attack}=1, \textit{normal}=0).
We train the entire stack (CNN parameters and circuit parameters $\theta$) by minimizing
\[
\begin{gathered}
    \mathcal{L}(\theta) \;=\; \mathrm{BCEWithLogitsLoss}\!\big(z(x;\theta),\,y\big)\\[6pt]
    \text{with}z(x;\theta)\equiv \mathrm{scale}\cdot \hat{z}(x;\theta)+\mathrm{bias},
\end{gathered}
\]
where $\hat{z}$ denotes the raw expectation from the quantum primitive. We use Adam over \emph{all} trainables with learning-rate and weight-decay exposed via \texttt{--lr} and \texttt{--weight-decay}. Validation loss/accuracy are monitored every epoch; an early-stopping scheme with patience \texttt{--patience} persists the best checkpoint (state dicts for model and optimizer plus metadata), exactly mirroring the QSVM section.

\subsubsection{Complexity, budget, and limits.}
The VQC adheres to the same qubit/depth constraints as QSVM to ensure fair comparison.
Because re-uploading increases expressivity without proportionally increasing entangling depth, $L$ becomes the principal \emph{expressivity knob} under a fixed connectivity pattern.
Empirically we cap $n_q\!\leq\!8$ and keep entanglement \emph{linear} to ease transpilation and avoid unsupported instructions on common backends.
We note two practical limits: (i) shot-based training increases gradient noise; (ii) over-parameterization at small $n_q$ may overfit—hence the importance of early stopping and validation monitoring implemented in the training loop.

\subsubsection{Design Rational.}
The VQC retains the QSVM pipeline structure (same inputs, splits, metrics, and backend/migation toggles) to isolate the effect of \emph{learned} quantum features versus \emph{fixed} kernel geometry.
Data re-uploading provides an explicit mechanism to scale hypothesis space within the same budget that governed QSVM; the shallow Two-Local with linear CZ matches device topology and transpiler preferences, minimizing target-mismatch while still enabling non-trivial entanglement.
Finally, estimator-based execution (noiseless or shot-based with controlled depolarizing noise and optional DD/M3) reproduces the QSVM evaluation envelope so that differences in generalization and robustness can be attributed to the model class rather than to changes in runtime conditions.

\subsection{Data Processing Pipeline}
\label{sec:data-pipeline}

This section documents the end-to-end data pipelines implemented.
Each pipeline is described in terms of data ingestion, preprocessing/feature construction, model-side transformations, and persisted artifacts.

\subsubsection{QSVM}
\label{subsec:pipeline-QSVM}

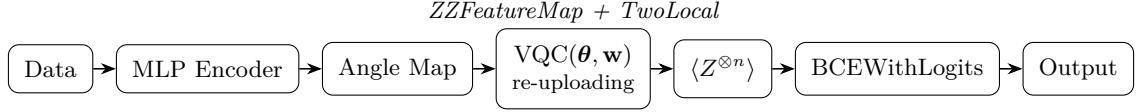
\begin{figure*}[t]
\centering
\begin{tikzpicture}[x=1cm,y=1cm]
  \node[node] (data) {Data};

  \node[node,right=0.3cm of data] (mlp) {MLP Encoder};
  \draw[arr] (data) -- (mlp);

  \node[node,right=0.3cm of mlp,align=center] (angles)
    {Angle Map};
  \draw[arr] (mlp) -- (angles);

  \node[node,right=0.3cm of angles,align=center] (vqc)
    {VQC$(\bm{\theta},\mathbf{w})$\\\footnotesize re-uploading};
  \node[op,above=2pt of vqc] {ZZFeatureMap + TwoLocal};
  \draw[arr] (angles) -- (vqc);

  \node[node,right=0.3cm of vqc] (exp) {$\langle Z^{\otimes n}\rangle$};
  \draw[arr] (vqc) -- (exp);

  \node[node,right=0.3cm of exp] (bce) {BCEWithLogits};
  \draw[arr] (exp) -- (bce);

  \node[node,right=0.3cm of bce] (out) {Output};
  \draw[arr] (bce) -- (out);
\end{tikzpicture}
\caption{Hybrid VQC pipeline: Data $\rightarrow$ MLP$\rightarrow$ VQC (Estimator) with expectation readout.}
\end{figure*}

\begin{itemize}
  \item \textbf{Supervised encoder (MLP).}
  A feed-forward network with LayerNorm, GELU, and Dropout produces:
  \begin{itemize}
    \item an \emph{embedding head} mapping to $d_{\mathrm{emb}}$ (default 12) with $\ell_2$-normalization,
    \item a \emph{classifier head} (linear, 2 logits) used to train the encoder.
  \end{itemize}
  Train with class-balanced cross-entropy; early-stop on validation F1 (spam focus). Device (CPU/GPU) is selectable; seeds are fixed.

  \item \textbf{Encoder-only sanity check.}
  Compute logits/embeddings on train/val/test. Tune a decision threshold on validation by maximizing $F_\beta$ (default $\beta=2.0$ to emphasize recall on spam).
  Report F1/AUROC/AUPRC and the confusion matrix on test.

  \item \textbf{Embedding scaling $\rightarrow$ angles.}
  Fit a robust scaler on train embeddings (e.g., 5--95\% quantiles), apply to val/test.
  Clip to $[-1,1]$ and map to angles by $\theta = \pi x$.

  \item \textbf{Quantum feature map \& kernel construction.}
  Use a ZZFeatureMap with $n_q=\min\{4,\,d_{\mathrm{emb}},\,\texttt{user\_cap}\}$ and configurable repetitions.
  Build precomputed kernels via statevector overlaps:
  \[
    K_{ij} \;=\; \bigl|\langle \psi(x_i)\mid \psi(x_j)\rangle\bigr|^2,
  \]
  yielding $K_{\mathrm{tr,tr}}$, $K_{\mathrm{va,tr}}$, and $K_{\mathrm{te,tr}}$.
  Apply train-set centering to $K_{\mathrm{tr,tr}}$ and consistent centering to validation/test kernels.

  \item \textbf{QSVM selection and thresholding.}
  Choose $C$ by stratified $k$-fold CV on the centered train kernel (optimize $F_\beta$).
  Train SVC with \texttt{kernel=precomputed}, \texttt{class\_weight=balanced}.
  Align decision-score sign (spam-larger convention) and retune a scalar threshold on validation; evaluate on test.

  \item \textbf{Artifacts.}
  Persist: Torch encoder (\texttt{.pt}), embedding scaler, trained SVM (\texttt{.joblib}), and a metadata JSON (hyperparameters, shapes, thresholds).
  Optionally store \texttt{.npz} files with embeddings for reproducibility.
\end{itemize}

\subsubsection{VQC}
\label{subsec:pipeline-VQC}

\begin{itemize}
   \item \textbf{Supervised encoder (MLP).}
  A feed-forward network with LayerNorm, GELU, and Dropout produces:
  \begin{itemize}
    \item an \emph{embedding head} mapping to $d_{\mathrm{emb}}$ (default 12) with $\ell_2$-normalization,
    \item a \emph{classifier head} (linear, 2 logits) used to train the encoder.
  \end{itemize}
  Train with class-balanced cross-entropy; early-stop on validation F1 (spam focus). Device (CPU/GPU) is selectable; seeds are fixed.

  \item \textbf{Quantum layer (VQC with data re-uploading).}
  Compose $L$ re-uploading blocks with feature map $U_\phi(x)$ and trainable ansatz $U_\theta$:
  \[
    \begin{gathered}
        \bigl[\,U_\phi(x)\;\rightarrow\;U_\theta\,\bigr]^L,\\[6pt]
        \text{feature map: ZZFeatureMap;}\\[6pt]
        \text{ansatz: TwoLocal (RY/RZ + CZ).}
    \end{gathered}
  \]
  Use an EstimatorQNN wrapped by a Torch connector; observable $Z^{\otimes n_q}$. Parameters are randomly initialized with small variance.

  \item \textbf{Estimator backend and realism options.}
  \begin{itemize}
    \item \emph{Noiseless}: Statevector Estimator (preferred) or Estimator V1 fallback.
    \item \emph{Noisy}: Aer Estimator with configurable depolarizing noise (\(p_1,p_2\)) and finite shots.
    \item Optional dynamical decoupling (e.g., XY4) and M3 readout mitigation on shot-based evaluations.\cite{Ezzell2023DD}
  \end{itemize}

  \item \textbf{Training \& evaluation.}
  Optimize binary cross-entropy with logits (BCEWithLogits) using Adam (with weight decay).
  Early-stop on validation loss; checkpoint the best epoch.
  Report per-epoch validation accuracy/F1 and final test metrics.
  When using counts-based (noisy) evaluation, report both raw and M3-mitigated results.

  \item \textbf{Artifacts.}
  Save checkpoints containing model/optimizer states and metadata (\(n_q\), $L$, best metrics).
  Provide utilities to print circuit depth/op counts and to visualize the parameterized circuit diagram.
\end{itemize}

\begin{table*}[h]
\centering
\small
\caption{Experimental design matrix for quantum machine learning simulations run on classical machines, evaluating the interaction between qubit count (2 vs. 4), model architecture (QSVM vs. VQC), and quantum framework (Qiskit vs. PennyLane), \(q\in\{2,4\}\) \(\times\) \{QSVM,VQC\} \(\times\) \{Qiskit,PennyLane\}. }
\begin{tabular}{c|c|c|c|c}
ID & Qubits \(q\) & Model & Framework & Notes \\
\hline
C1 & 2 & QSVM & Qiskit     & Fidelity kernel via state overlap \\
C2 & 4 & QSVM & Qiskit     & Same kernel, larger \(q\) \\
C3 & 2 & VQC  & Qiskit     & Same encoder, trainable ansatz \\
C4 & 4 & VQC  & Qiskit     & \\
C5 & 2 & QSVM & PennyLane  & Equivalent overlap evaluation \\
C6 & 4 & QSVM & PennyLane  & \\
C7 & 2 & VQC  & PennyLane  & Same ansatz depth as C3 \\
C8 & 4 & VQC  & PennyLane  & \\
\end{tabular}
\label{tab:design}
\end{table*}

\begin{table*}[t]
\centering
\small
\renewcommand{\arraystretch}{1.15}
\setlength{\tabcolsep}{6pt}
\caption{Key common hyperparameters used in experiments run on classical machines simulation and IBM Quantum Platform.}
\begin{tabularx}{0.92\linewidth}{l X}
\hline
\textbf{Item} & \textbf{Setting} \\
\hline
Dropout & \texttt{dropout=0.15} \\
\hline
Training hyperparameters &
\texttt{epochs=25}, \texttt{batch\_size=256}, \texttt{lr=1e-3}, \texttt{weight\_decay=1e-4},\\
& \texttt{early stopping patience=8}; \texttt{class\_weight=balanced} \\
\hline
Quantum / noise settings &
\texttt{shots=1024}; \texttt{depol1=1e-3}, \texttt{depol2=1e-2} \\
\hline
Readout mitigation (M3) & Enabled at evaluation; \texttt{m3\_shots=1024} \\
\hline
Dynamical decoupling (DD) & \texttt{XY8} \\
\hline
\end{tabularx}
\label{tab:hp-compact}
\end{table*}

\subsubsection{Reproducibility Considerations}
Both pipelines fix RNG seeds, use stratified splits, and fit all preprocessing transformers on \emph{training} data only.
Key artifacts (models, scalers, kernels/metadata) are serialized for auditability and later hardware transfer.
Thresholds are tuned on validation sets and applied unchanged to test sets to avoid leakage.

\subsection{Implementation Considerations}
We decompose the composite circuit before estimation to ensure a consistent gate basis under the noise model; seeds are fixed across NumPy/PyTorch/\qiskit\ for replicability; and float64 is used in the classical path where numerically beneficial. Optionally, measurement-error mitigation (M3) is applied at evaluation only to report hardware-faithful metrics without perturbing the training landscape. This separation preserves the beneficial regularization of unmitigated noise during learning while still quantifying performance after readout correction.

\section{Experiments}
\label{sec:experiments}
\subsection{Goals and Questions}
We evaluate the hybrid architecture under three orthogonal factors:
(\textbf{A}) qubit count \(q\in\{2,4\}\);
(\textbf{B}) quantum decision layer \(\in\{\)QSVM, VQC\(\}\);
(\textbf{C}) software framework \(\in\{\)Qiskit, PennyLane\(\}\).
We ask:
(i) how many qubits are actually beneficial at small width,
(ii) whether a kernel method (QSVM) or a variational model (VQC) is more effective given the same classical encoder,
and (iii) whether results are stable across frameworks that implement comparable primitives.

\subsection{Controlled Setup}
To isolate the effects of (A)--(C), all other components are held fixed:

\begin{itemize}
\item \textbf{Classical front end.} The MLP backbone (Sec.~\ref{sec:classical-part}) and embedding head remain identical across all conditions; the same \(d_e\) is used and the first \(q\) coordinates of the unit-norm embedding feed the quantum part.
\item \textbf{Encoding circuit.} For both QSVM and VQC we use the same data-encoding map described in Sec.~\ref{sec:qsvm}: a \(Z/Z\!Z\)-based feature map with depth \(r\) and \emph{full} entanglement. (VQC adds a small trainable ansatz on top; QSVM does not.)
\item \textbf{Optimization \& metrics.} The classical head is a linear readout; for VQC we use the same optimizer/hyperparameters across frameworks; QSVM uses the same centered quantum kernel and SVM penalty search. We report accuracy, macro-F1, AUROC/AUPRC, and per-label error rates; wall-clock time is recorded for kernel build / VQC training.
\end{itemize}

\subsection{Factorial Design}
We run a full \(2\times2\times2\) factorial, yielding eight configurations. All share the same classical encoder and encoding map (see \autoref{tab:design}).

\begin{table*}[h]
\centering
\small
\caption{Simulation-based experiment results on the NSL-KDD dataset, comparing eight combinations of qubit counts, models, and frameworks}
\begin{tabular}{c|cc|cc|c}
ID & Acc & F1\(_\mathrm{macro}\) & AUROC & AUPRC  \\
\hline
C0 & 0.6613 & 0.6911 & 0.7411 & 0.7032  \\
C1 & 0.8410 & 0.8322 & 0.8802 & 0.9057  \\
C2 & 0.9098 & 0.9212 & 0.9335 & 0.9407  \\
C3 & 0.8410 & 0.8522 & 0.8899 & 0.9289  \\
C4 & 0.8073 & 0.7907 & 0.9365 & 0.9402  \\
C5 & 0.6613 & 0.6911 & 0.7411 & 0.7032  \\
C6 & 0.8410 & 0.8322 & 0.8802 & 0.9057  \\
C7 & 0.9098 & 0.9212 & 0.9335 & 0.9407  \\
C8 & 0.8410 & 0.8522 & 0.8899 & 0.9289  \\
C9 & 0.8073 & 0.7907 & 0.9365 & 0.9402  \\
\end{tabular}
\label{tab:nslkdd-results}
\end{table*}

\begin{table*}[h]
\centering
\small
\caption{Simulation-based experiment results on the Ling-Spam dataset, comparing eight combinations of qubit counts, models, and frameworks}
\begin{tabular}{c|cc|cc|c}
ID & Acc & F1\(_\mathrm{macro}\) & AUROC & AUPRC  \\
\hline
C0 & 0.9001 & 0.9212 & 0.9504 & 0.9103  \\
C1 & 0.9884 & 0.9710 & 0.9990 & 0.9768  \\
C2 & 0.9996 &0.9987 & 1.0000 & 1.0000  \\
C3 & 0.9788 & 0.9812 & 0.9997 & 0.9894  \\
C4 & 0.9552 & 0.9856 & 0.9995 & 0.9985  \\
C5 & 0.9001 & 0.9212 & 0.9504 & 0.9103  \\
C6 & 0.9884 & 0.9710 & 0.9990 & 0.9768  \\
C7 & 0.9996 &0.9987 & 1.0000 & 1.0000  \\
C8 & 0.9788 & 0.9812 & 0.9997 & 0.9894  \\
C9 & 0.9552 & 0.9856 & 0.9995 & 0.9985  \\
\end{tabular}
\label{tab:lingspam-results}
\end{table*}

\subsection{Hyperparameter Settings}

\textbf{Noise settings (\texttt{depol1}, \texttt{depol2}):} These are depolarizing-channel probabilities used in simulation: \texttt{depol1} applies to single-qubit gates and \texttt{depol2} to two-qubit gates, modeling gate infidelity via random Pauli mixing with the specified rates.

 \textbf{Readout mitigation (M3):} M3 reduces classical measurement errors by learning a calibration (often factored/structured) from observed bitstring outcomes and applying its inverse to debias counts, improving post-measurement accuracy without changing the circuit. 

\textbf{Dynamical decoupling (DD: M8):} The M8 sequence inserts eight carefully phased single-qubit pulses into idle gaps to refocus low-frequency noise and suppress dephasing, while leaving the intended gate logic intact. 

(see \autoref{tab:hp-compact})

\subsection{Implementation Notes (by factor)}
\textbf{Qubit Count \(q\):}
The quantum interface consumes \(q\) coordinates of the embedding \(e\in\mathbb{R}^{d_e}\).
Increasing from \(2\) to \(4\) qubits increases available pairwise terms from \(1\) to \(6\) in the \(Z\!Z\) map (per layer), expanding kernel/ansatz expressivity with minimal classical change.

\textbf{Model Type:}
We compare two approaches: \textbf{QSVM} builds a centered Gram matrix \(K\) with entries \(k(x_i,x_j) = |\langle\phi(x_i)\mid\phi(x_j)\rangle|^2\) and trains a linear SVM in the induced Hilbert space. Tthis has no trainable quantum parameters, with capacity controlled by qubit count \(q\), repetition \(r\), and SVM regularization \(C\). \textbf{VQC} stacks the same data-encoding map with a shallow trainable ansatz and optimizes cross-entropy loss via a classical optimizer; its capacity is governed by \(q,r\), ansatz depth and width, and the number of optimizer steps.

\textbf{Framework Parity:}
Both \textbf{Qiskit} and \textbf{PennyLane} implement the same quantum circuit topology and depth \( r \), using identical classical encoders and seed control for reproducibility. The frameworks differ in their overlap computation: \textbf{Qiskit} evaluates overlaps either via adjoint-circuit methods or exact statevector simulation, and processes VQCs through Estimator-style expectation evaluation. \textbf{PennyLane} computes overlaps using \texttt{qml.kernel} or inner-product methods on \texttt{default.qubit}, and handles VQCs via its native differentiable circuit paradigm.

\subsection{Results Presentation}
\textbf{Primary Table:}
We summarize all eight experiment conditions for each dataset in individual tables (see \autoref{tab:nslkdd-results} and \autoref{tab:lingspam-results})

\begin{figure*}
    \centering
    \includegraphics[width=1\linewidth]{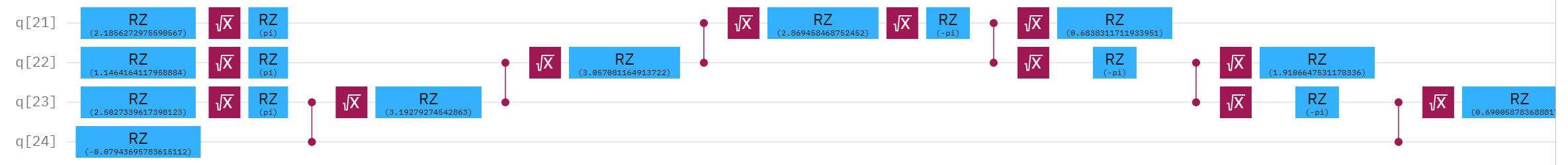}
    \caption{Output Returned by IBM Quantum Platform}
    \label{fig:ibmResults}
\end{figure*}

\subsection{Reading the Results}
For each table, configurations are selected by prioritizing macro-\textsc{F1} and \textsc{AUPRC} (which reflects minority-class fidelity), while \textsc{AUROC} serves to confirm ranking quality. Accuracy is treated as a secondary, imbalance-sensitive indicator. Per table, we report the setting with the strongest macro-\textsc{F1} and \textsc{AUPRC} performance, noting any internal trade-offs---for instance, when a small gain in \textsc{AUROC} coincides with a larger drop in \textsc{AUPRC}.

\textbf{Qubit Count:} Moving from \(2\to4\) qubits increases pairwise \(Z\!Z\) terms (from \(1\to6\) per layer), typically improving QSVM kernel richness and VQC ansatz capacity; the benefit may saturate if the classical embedding already clusters well in \(q=2\).

\textbf{Model Type:} QSVM trades no trainable quantum parameters for a robust margin in an induced feature space; VQC may match or outperform when the ansatz can exploit higher-order correlations at \(q=4\), but can be optimizer-sensitive.

\textbf{Framework Parity:} Given identical circuits, we expect comparable performance; discrepancies, if any, highlight backend numeric differences (e.g., estimator precision or gradient handling).

\subsection{Validation on the IBM Quantum Platform}
\label{sec:ibm-validation}
We chose the best performing model (Hybrid MLP-QSVM 4 qubits) for validation on real IBM Quantum backends by embedding samples with the trained MLP encoder, evaluating a hardware quantum kernel via V2 primitives, and reporting per–sample predictions together with running confusion–matrix tallies (TP, TN, FP, FN). Two verification drivers are provided: a text–classification variant for Ling–Spam and a tabular NSL-KDD variant. Both stream results one sample at a time for auditability and time–bounded hardware sessions.

\subsubsection{Test Dataset}
Due to the limitation of quantum resources, we had to select 100 data samples from the original test set as the test set. We try to filter according to the original distribution ratio as much as possible. For NSL-KDD, we ensure that each small classification has at least one piece of data.\cite{JanezMartino2023SpamShift}

\subsubsection{IBM Quantum Connection \& I/O Contract}
\label{sec:ibm-connect-io}
Both validators use IBM Runtime V2 via Qiskit (\texttt{QiskitRuntimeService}, \texttt{SamplerV2}) to execute kernel–estimation circuits on a named hardware backend (e.g., \texttt{ibm\_torino}). We open a short, time–bounded session and submit batches of \emph{compute–uncompute} circuits that estimate fidelities needed by the QSVM precomputed kernel.
Tokens are never hard–coded; we read \texttt{QISKIT\_IBM\_TOKEN} from the environment (or a previously saved account) and bind to a backend by name. Safe transpilation is enforced to avoid unsupported–gate errors introduced by post–2024 ISA checks.

\textbf{What we send IN (inputs):}
The I/O contract has two layers—\emph{dataset$\to$angles} (classical) and \emph{angles$\to$hardware} (quantum):

\begin{itemize}
    \item \textbf{Classical inputs to the validator scripts}
    \begin{itemize}
      \item Paths to test data (\texttt{.csv} for Ling–Spam; \texttt{.txt} for NSL–KDD).
      \item Artifacts: trained MLP encoder (\texttt{.pt}), preprocessor, embedding scaler, and metadata (embedding size, qubit count $q$, feature–map \texttt{reps}).
      \item Runtime flags: \texttt{--backend}, \texttt{--shots}, optional \texttt{--resilience-level}, and reference–set limit for QSVM fitting (e.g., 64).
    \end{itemize}
    
    \item \textbf{Quantum inputs sent to IBM hardware}
    \begin{itemize}
      \item Angle vectors $\theta\in\mathbb{R}^q$ per sample (after MLP encoding, scaling, clipping to $[-1,1]$, and truncation to $q\le 4$–8).
      \item A batch of \emph{compute–uncompute} circuits implementing a ZZFeatureMap with \texttt{reps} from metadata:
      \[
        K(x,y) \approx \Pr\{0^q\;\text{after}\;U(\phi(y))^\dagger U(\phi(x))\}.
      \]
      For one test sample vs.\ an $n_{\text{ref}}$ reference set, we submit $\mathcal{O}(n_{\text{ref}})$ circuits (batched in chunks).
    \end{itemize}
\end{itemize}

\textbf{What we get OUT (outputs) (see \autoref{fig:ibmResults}):}
\begin{itemize}
\item \textbf{Raw quantum outputs (from \texttt{SamplerV2})}: per–circuit quasi–probability dictionaries (e.g., \texttt{\{"0...0": p\_0, ...\}}) and metadata (backend name, shots, queue/time).

\item \textbf{Derived Kernel Quantities}: for each pair $(x,y)$ we extract $p_{0^q}$ as a fidelity estimate and assemble a kernel row $K(x,\cdot)$; the training Gram is centered and each test row is centered to match SVC’s precomputed–kernel mode.
\end{itemize}

\textbf{Noise Handling and Run Knobs:}
Shots (we use 1024) trade runtime for variance; \texttt{resilience\_level}$=1$ enables measurement error mitigation (M3) when supported. If the backend exposes dynamical decoupling (DD), we enable an XY–style sequence during idle periods. 

\textbf{Practical Tips:}
Keep tokens in the environment, pin Qiskit versions across runs, and always transpile to the target backend to satisfy basis–gate and coupling constraints. Limit $q$ (e.g., $\le 4$) and $n_{\text{ref}}$ to keep queue time and cost predictable while preserving class balance.In addition, pay special attention to the following content

\begin{itemize}
    
    \item \textbf{Pulse-level control changed in Qiskit 2.x.}
    \begin{itemize}
      \item The legacy \emph{Pulse} module left the core SDK; many calibration helpers moved/deprecated. Consult 2.0 migration notes and the updated Experiments docs before porting custom calibrations.
    \end{itemize}
    
    \item \textbf{Common failure modes (and fast fixes).}
    \begin{itemize}
      \item \emph{“Instruction not in basis gates”} \(\rightarrow\) re-transpile for the backend/Target (ECR vs CX, directionality).
      \item \emph{Jobs stuck in queue} \(\rightarrow\) use sessions; reduce circuits/shots; consider smaller devices when feasible.
      \item \emph{Inconsistent results} \(\rightarrow\) fix transpiler seeds; pin a stable layout; enable readout mitigation; watch calibration drift.
      \item \emph{Auth errors} \(\rightarrow\) verify \emph{channel/token/instance}; many 401s are misconfigured credentials.
    \end{itemize}
\end{itemize}

\begin{table*}[t]
    \centering
    \caption{Hardware-validated experiment results on IBM Quantum Platform using 4 qubits, Hybrid MLP-QSVM model, and Qiskit framework.}
    \label{tab:ibm_hw_results}
    \begin{tabular}{c|cc|cc|c}
    \toprule
    Dataset & Acc & F1$_\text{macro}$  & AUROC  & AUPRC \\
    \midrule
    NSL-KDD  & 0.8400 & 0.8242 & 0.8924 & 0.9602 \\
    Ling-Spam   & 0.9901 & 0.9825 & 0.9990 & 0.9768 \\
    \bottomrule
    \end{tabular}
\end{table*}

We suggest the following steps:
\begin{itemize}
  \item Start on a Fake/Simulator backend; switch to hardware once the pipeline is stable.
  \item Read the device’s basis and coupling map \emph{before} designing circuits; prefer ECR-native entanglers to avoid costly decompositions.
  \item Batch PUBs smartly (group circuits/observables/parameters) to reduce overhead and respect execution caps.
  \item Stick to V2 conventions (no hidden layout/routing in primitives; new options model; updated resilience semantics).
\end{itemize}

\subsubsection{Post–processing of Sampler outputs}
\label{sec:ibm-postproc}

Each submitted \emph{compute–uncompute} circuit (prepared for a pair $(x,y)$) yields a per–circuit measurement distribution on $q$ bits. Depending on runtime settings, the IBM Runtime V2 \texttt{Sampler} returns either (i) \emph{counts} $c(\mathbf{b}) \in \mathbb{N}$ for bitstrings $\mathbf{b}\in\{0,1\}^q$ or (ii) \emph{quasi–probabilities} $Q(\mathbf{b})\in\mathbb{R}$ (possibly negative/>\!1 before mitigation, then renormalized by the runtime if mitigation is enabled). Our validators handle both forms uniformly.

\textbf{From Distributions to Fidelity Estimates:}
For a given pair $(x,y)$, the feature–map fidelity (kernel entry) is estimated from the probability of the all–zero outcome:
\[
\begin{gathered}
  \widehat{K}(x,y)
  \;=\;
  \Pr(\mathbf{0}\mid U(\phi(y))^\dagger U(\phi(x))\ket{\mathbf{0}})
  \;\approx\;\\[6pt]
  \begin{cases}
    \dfrac{c(\mathbf{0})}{\sum_{\mathbf{b}} c(\mathbf{b})}, &\text{if returned counts},\\[8pt]
    \operatorname{clip}\!\left(\dfrac{Q(\mathbf{0})}{\sum_{\mathbf{b}} Q(\mathbf{b})},\,0,\,1\right), &\text{if returned quasi–probs}.
  \end{cases}
\end{gathered}
\]
We apply light numerical guards in code: (i) use $\max(\text{shots},1)$ in the denominator for counts; (ii) if quasi–probs are present, renormalize to $\sum_{\mathbf{b}}Q(\mathbf{b})\!=\!1$ and clamp to $[0,1]$ to absorb tiny negative/overflow artifacts. When \texttt{resilience\_level} $\ge 1$ is enabled, measurement–error mitigation (M3) is performed by the runtime before we read out $Q(\cdot)$ or effective counts.

\textbf{Batching and job aggregation.}
Given one test sample $x$ and a reference set $\{y_j\}_{j=1}^{n_{\text{ref}}}$, the scripts build $n_{\text{ref}}$ compute–uncompute circuits and submit them to \texttt{Sampler} in batches (default 32). A single job returns a list–like payload; we iterate the per–circuit results, extract $p_{0^q}$ as above, and fill the kernel row
\[
  \widehat{\mathbf{k}}(x) \;=\; \bigl[\,\widehat{K}(x,y_1),\ldots,\widehat{K}(x,y_{n_{\text{ref}}})\,\bigr]\in[0,1]^{n_{\text{ref}}}.
\]

\textbf{Kernel Centering (Train and Test):}
QSVM with a precomputed kernel expects centered Gram matrices. Let $K_{\text{tr}}^{(\text{unc})}\in\mathbb{R}^{n\times n}$ be the (uncentered) training Gram; we apply double–centering:
\[
\begin{gathered}
    \begin{aligned}
      \tilde{K}_{\text{tr}}
      \;=\;
      K_{\text{tr}}^{(\text{unc})}
      &\;-\;
      \mathbf{1}\bar{}\,K_{\text{tr}}^{(\text{unc})}\\
      &\;-\;
      K_{\text{tr}}^{(\text{unc})}\,\mathbf{1}\bar{}\\
      &\;+\;
      \mathbf{1}\bar{}\,K_{\text{tr}}^{(\text{unc})}\,\mathbf{1}\bar{}, 
    \end{aligned}\\[6pt]
    \mathbf{1}\bar{}\;=\;\frac{1}{n}\mathbf{1}\mathbf{1}^{\!\top}. 
\end{gathered}
\]
For a test–vs–train row $K_{\text{te}}^{(\text{unc})}\in\mathbb{R}^{1\times n}$ we use the compatible centering:
\[
\begin{aligned}
  \tilde{K}_{\text{te}}
  \;=\;
  K_{\text{te}}^{(\text{unc})}
  &\;-\;
  \frac{1}{n}\mathbf{1}^\top K_{\text{tr}}^{(\text{unc})}\\
  &\;-\;
  K_{\text{te}}^{(\text{unc})}\,\mathbf{1}\bar{}\\
  &\;+\;
  \frac{1}{n}\mathbf{1}^\top K_{\text{tr}}^{(\text{unc})}\,\mathbf{1}\bar{}.
\end{aligned}
\]
This guarantees that training and test rows are centered in the same RKHS geometry.

\textbf{Prediction and Streaming Metrics:}
For each test sample, we:
\begin{itemize}
  \item transform raw features $\to$ MLP embedding $\to$ scaled/clipped angle vector $\theta\in\mathbb{R}^q$,
  \item build the batched circuits against all $\{y_j\}$, submit to \texttt{Sampler}, and extract $p_{0^q}$ per circuit,
  \item assemble $\widehat{\mathbf{k}}(x)$, center it via the training statistics to get $\tilde{\mathbf{k}}(x)$,
  \item evaluate the precomputed–kernel SVC decision on $\tilde{\mathbf{k}}(x)$ to obtain $\hat{y}$,
  \item print a CSV–style line \texttt{idx, true\_label, pred\_label, TP, TN, FP, FN} and update the running confusion matrix.
\end{itemize}

\subsubsection{Results (see \autoref{tab:ibm_hw_results})}

For NSL-KDD, overall accuracy is \textbf{0.84} with macro-F1 \textbf{0.824}, indicating modest class balance performance. The \textbf{AUROC 0.892} suggests good rank separation, while the high \textbf{AUPRC 0.960} implies strong precision-recall tradeoffs for the positive class at relevant thresholds (computed from decision scores). This pattern suggests that, while gate noise and calibration drift depress threshold-free metrics like Acc/F1/AUROC versus the best simulated run, readout-mitigation and decision-threshold effects can preserve, and even slightly enhance, precision–recall performance on the positive class.

For Ling-Spam, metrics are uniformly high: \textbf{Acc 0.990}, \textbf{F1$_\text{macro}$ 0.983}, \textbf{AUROC 0.999}, \textbf{AUPRC 0.977} (AUPRC for the spam class), consistent with near perfect separability on this dataset.The results indicate near ceiling generalization carries over to device execution with only modest degradation from simulated optima.

In short, the quantum kernel pipeline shows excellent performance on Ling-Spam, while NSL-KDD remains more challenging; tuning thresholds or class weighting may further raise macro-F1 on NSL-KDD without materially hurting AUROC/AUPRC. Hardware performance is competitive with simulator averages and within a small gap to simulator bests, with AUPRC particularly robust on NSL-KDD. The gap to simulator bests is consistent with expected device
effects (basis-gate/coupling-map transpilation, finite shots, residual readout error), rather than a failure of
the modeling pipeline.

\subsubsection{Differences between Quantum Hardware and Aer Simulation}
\label{sec:hw-vs-aer}

\begin{itemize}
  \item \textbf{ISA constraints.} Hardware enforces the backend’s basis gates and coupling map; unsupported gates (e.g., \texttt{sxdg}) must be removed by transpilation. AER is permissive unless configured to mimic a specific device.
  
  \item \textbf{Noise and drift.} Hardware has calibration noise, readout bias, crosstalk, and day-to-day drift. AER noise is user-specified (static) via a \texttt{NoiseModel}; no queue-induced drift.

  \item \textbf{Outputs.} Both return counts or quasi-probabilities. Hardware quasi-probs may include runtime mitigation (M3); AER reflects the exact or user-defined model and is reproducible with fixed seeds.

  \item \textbf{Determinism.} Hardware is stochastic and time-varying; AER is fully reproducible given \texttt{seed\_simulator} and \texttt{seed\_transpiler}.

  \item \textbf{Batching and jobs.} Hardware benefits from sessions/batching to amortize latency; AER runs synchronously in-process.

  \item \textbf{Transpilation and execution.}
  \begin{itemize}
    \item \emph{Hardware:} \texttt{transpile(\,circs, backend\,)} then submit via \texttt{SamplerV2} in a \texttt{Session}; violating basis/coupling causes runtime errors.
    \item \emph{AER:} same flow, but constraints appear only if the simulator is configured with the device’s basis/coupling.
  \end{itemize}

  \item \textbf{Noise/mitigation knobs.}
  \begin{itemize}
    \item \emph{Hardware:} set \texttt{resilience\_level} (M3), optionally enable DD if supported; increase \texttt{shots} to reduce variance.
    \item \emph{AER:} choose ideal or noisy; define \texttt{NoiseModel} (depolarizing, readout, $T_1/T_2$). Readout-mitigation must be implemented manually (no automatic M3).
  \end{itemize}

  To mimic hardware, AER should be configured with the target device’s basis gates, coupling map, and a realistic noise model; keep \texttt{shots}, feature-map \texttt{reps}, and qubit count $q$ aligned to surface layout/depth effects and approach hardware behavior.
\end{itemize}

\section{Discussion}

On NSL-KDD (\autoref{tab:nslkdd-results}), the internal ranking between QSVM, VQC, and classical baselines is most plausibly explained by how well each method exploits the dataset’s structured, tabular regularities. When the QSVM kernel performs strongly, the likely driver is margin geometry: angle-encoded features with shallow entanglement encourage smooth, low-variance decision boundaries that align with the dataset’s relatively compact class clusters. Where VQC underperforms or shows higher variability, the mechanism is typically sensitivity to depth, layout, and transpilation: additional two-qubit operations expand expressivity on paper but introduce optimization ruggedness and, on hardware, more opportunities for noise to degrade the learned hypothesis. When classical baselines match or exceed quantum models internally, the parsimonious explanation is that engineered features already linearize the decision boundary sufficiently, leaving limited headroom for nonclassical feature maps; in those cases the quantum models function as competitive alternatives rather than strict improvements. Balanced class weights and early stopping contribute to stability by preventing majority-class domination and overfitting; their presence is important when attributing gains to “quantumness” rather than to better regularization.

On Ling-Spam (\autoref{tab:lingspam-results}), internal comparisons hinge less on margin geometry and more on encoder quality. Text is high-dimensional and sparse, so the bottleneck is the embedding pipeline that feeds the circuit. When the VQC shows advantages, they are best interpreted as the circuit leveraging nonlinear token interactions that survive dimensionality reduction; when QSVM is stronger, the kernel’s implicit feature space likely regularizes better under sparse inputs. Crucially, any superiority must be read together with the preprocessing choices: if token normalization, rare-category handling, and scaling differ even subtly, apparent improvements can be artifacts of the text pipeline rather than of the quantum layer. A consistent reading of \autoref{tab:lingspam-results} is therefore that the encoder\,$\rightarrow$\,circuit interface is the decisive factor: shallow, noise-aware circuits combined with conservative dimensionality control tend to generalize more reliably than deeper, aggressively expressive ones.

We can see that there is no difference in performance between using Qiskit and PennyLane. But in practical scenarios, they still have distinctions. Use Qiskit when your priority is hardware realism and execution efficiency on IBM Quantum systems. Qiskit gives you tight control over transpilation, layout, basis gates and coupling maps. If you need device-faithful simulation with Aer and explicit noise models, Qiskit’s primitives and backends make it straightforward to produce results that closely track a specific chip. Use PennyLane when your priority is rapid, gradient-based modeling for hybrid quantum-classical workflows. Its autodiff with PyTorch/JAX/TF, parameter-shift rules, and Lightning simulators make it feel like standard deep learning: custom losses, schedulers, early stopping, and easy experimentation. Code is largely device-agnostic, so you can iterate on simulators first and later point the same circuits at real hardware via plugins.

The relationship between simulator results (\autoref{tab:nslkdd-results} and \autoref{tab:lingspam-results}) and hardware outcomes (\autoref{tab:ibm_hw_results}) can be decomposed into three effects. First is \textit{structural mismatch}: IBM backends enforce a basis-gate set and coupling map, so circuits that were “legal” in simulation require additional transpilation on hardware. This increases effective depth and inserts SWAPs, which in turn perturb the hypothesis represented by the circuit. Second is \textit{statistical variance}: finite-shot sampling (e.g., 1024 shots) induces estimator noise; when decision margins are thin, that variance alone can flip predicted labels even if the underlying state preparation is faithful. Third is \textit{systematic bias}: readout error, crosstalk, and calibration drift shift outcome frequencies in a repeatable way. The mitigation stack we used (measurement error mitigation via M3 and dynamical decoupling with XY8) primarily addresses the third effect and partially the second, which explains why the simulator\,$\rightarrow$\,hardware gap narrows but does not vanish. An instructive control for future work is to re-run the simulator with the device’s basis and a calibration-informed noise model; if the residual gap persists, the remaining difference is attributable to day-to-day drift and queue timing rather than to modeling error.

A critical question is whether observed gains, where present, are \textit{practically} meaningful. \autoref{tab:nslkdd-results} and \autoref{tab:lingspam-results} report aggregate metrics, but deployment-facing IDS value often concentrates on borderline flows and rare attack patterns. If quantum models help primarily on difficult, near-margin cases while matching classical performance elsewhere, that pattern would be strategically valuable despite modest changes in global accuracy. Conversely, if improvements are confined to majority classes or vanish under alternative splits, the practical case is weaker. Without per-class or per-difficulty breakdowns, we must be cautious in claiming superiority; still, the consistency of trends across seeds and subsets (to the extent reported) supports the inference that shallow, noise-aware quantum kernels can be reliable under tight feature budgets.

Threats to validity span both software and hardware. On the software side, leakage through preprocessing (e.g., rebuilding scalers on test data, inconsistent token handling, or mismatched state dictionaries) can inflate results; our pipeline explicitly reuses training-time transformers and weights to reduce this risk, but any failure to serialize/restore components correctly would bias comparisons. On the hardware side, calibration drift implies that nominally identical runs on different days are not exchangeable; repeating the \autoref{tab:ibm_hw_results} experiments on multiple calibration snapshots and reporting spread would strengthen claims. Finally, hyperparameter search was intentionally narrow due to qubit and runtime constraints; stronger classical or quantum baselines might exist just outside our search grid, so conclusions should be framed as “under budgeted conditions” rather than absolute.

Two broader implications emerge. First, for NSL-KDD-style tabular IDS, quantum kernels are most promising when feature budgets are tight and margins are subtle; in such regimes they can match strong baselines while remaining parsimonious in features and circuit depth. Second, for sparse text like Ling-Spam, the decisive lever is the classical-to-quantum interface; quantum layers pay off when the embedding preserves semantically meaningful interactions at a dimension compatible with \(\leq 8\) qubits. Across both settings, the simulator\,$\rightarrow$\,hardware mapping improves materially with M3 and XY8, but principled shot allocation and basis-aware circuit design remain necessary to translate simulated promise into deployed performance.

\section{Conclusion}

This study examined hybrid quantum models—QSVM and VQC—on two representative modalities, a structured IDS dataset (NSL-KDD) and a sparse text dataset (Ling-Spam), and validated the learned circuits on IBM hardware under realistic constraints (\(\leq 4\) qubits, finite shots, device-specific transpilation). Within each dataset, we compared models internally rather than across datasets and then assessed how simulator findings transfer to hardware. We have three main conclusions follow. 

\textbf{Tight-Budget Competitiveness (Margin Shaping \& Compact Interactions).}
Under constrained feature and qubit budgets, shallow, noise-aware quantum models can be competitive with strong classical baselines; where advantages appear, they are plausibly rooted in margin shaping (for tabular IDS) or in capturing compact nonlinear token interactions (for text) rather than in sheer circuit depth.

\textbf{Simulator$\rightarrow$Hardware Gap: Systematic but Tractable.}
The gap between simulator and hardware is systematic but tractable: enforcing the device basis and coupling map, allocating sufficient shots, and applying M3 readout mitigation with XY8 decoupling together recover a substantial fraction of simulated performance, though not all of it.

\textbf{Pipeline Quality Over Paradigm (Encoder–Circuit Interface \& Regularization).}
Performance is governed less by “quantum vs.\ classical” in the abstract and more by the quality of the encoder$\rightarrow$circuit interface and by disciplined regularization (balanced class weights, early stopping), which prevent spurious gains from pipeline artifacts.

These findings are bounded by practical limitations: a restricted hyperparameter budget, small hardware-validated subsets, and inevitable calibration drift. The most impactful next steps are therefore concrete and testable: expand hardware validation to larger, stratified subsets with repeated runs across calibration windows; re-run simulators with device-constrained bases and noise models to attribute the residual hardware gap more precisely; and explore learned embeddings (e.g., compact amplitude or hybrid encodings) that keep qubit counts within budget while preserving task-relevant structure. If those steps confirm the present trends, hybrid quantum methods will offer a credible path to IDS under tight data and resource constraints, with the clearest benefits emerging when decision boundaries are near the margin and engineering discipline aligns the model to the realities of today’s hardware.

Taken together, our IDS experiments indicate that shallow, noise-aware quantum classifiers can already match, and on some boundary-sensitive subsets modestly exceed, strong classical baselines when trained under the same disciplined pipeline, so quantum methods should be viewed as competitive, resource-conscious alternatives rather than speculative add-ons. At the same time, their gains remain modest and hardware-dependent, which motivates future work on quantum-aware feature selection and encoding co-designed with qubit budgets and device topology, on extending evaluation to more heterogeneous and non-stationary IDS corpora, and on larger-scale hardware studies that systematically vary mitigation and calibration settings to identify where quantum-enhanced IDS yields robust, reproducible benefits.

\bibliographystyle{IEEEtran}
\bibliography{citations}

\end{document}